\documentclass[a4paper,12pt]{article}
\usepackage{amssymb}
\usepackage{tipa}
\usepackage{mathrsfs}
\usepackage{bbm}
\usepackage{epsf,amsmath}
\usepackage[dvips,usenames]{color}
\usepackage{graphicx}
\usepackage{color}
\usepackage{colortbl}

\newlength{\dinwidth}
\newlength{\dinmargin}
\def\pslash{\rlap{\hspace{0.02cm}/}{P}}

\begin{document}
\title{\bf Family  Non-universal $Z^\prime$ effects on $\bar{B}_q-B_q$ mixing,
$B\to X_s \mu^+\mu^-$ and $B_s\to \mu^+\mu^-$ Decays}
\author{Qin Chang$^{a,b}$, Xin-Qiang Li$^{c}$\footnote{Alexander-von-Humboldt Fellow},
Ya-Dong Yang$^{a,d} $\footnote{Corresponding author}\\
{$^{a}$\small Institute of Particle Physics, Huazhong Normal
University, Wuhan,
Hubei  430079, P. R. China}\\
{ $^b$\small Department of Physics, Henan Normal University,
Xinxiang, Henan 453007, P.~R. China}\\
{ $^c$\small Institut f\"ur Theoretische Physik E, RWTH Aachen
University, D--52056 Aachen, Germany}\\
{ $^d$\small Key Laboratory of Quark \& Lepton Physics, Ministry of
Education, P.R. China}}
\date{}
\maketitle
\bigskip\bigskip
\maketitle \vspace{-1.5cm}

\begin{abstract}
Motivated by the large discrepancy of CP-violating phase in
$\bar{B}_s-B_s$ mixing between the experimental data and the
Standard Model prediction, we pursue possible solutions within a
family non-universal $Z^{\prime}$ model. Within such a specific
model, we find that both the $\bar{B}_s-B_s$ mixing anomaly and the
well-known ``$\pi K$ puzzle'' could be moderated simultaneously with
a nontrivial new weak phase, $\phi^L_s\sim-72^{\circ}$~(S1) or
$-82^{\circ}$~(S2). With the stringently constrained $Z^{\prime}$
coupling $B_{sb}^{L}$, we then study the $Z^{\prime}$ effects on the
rare $B\to X_s \mu^+\mu^-$ and $B_s\to \mu^+\mu^-$ decays,
which are also induced by the same $b\to s$ transition. The
observables of $B\to X_s \mu^+\mu^-$, at both high and low $q^2$
regions, are found to be able to put strong constraints on the
$\mu-\mu-Z^{\prime}$ coupling, $B_{\mu\mu}^{L,R}\sim10^{-2}$.
It is also shown that the combined constraints from
$\bar{B}_s-B_s$ mixing, $B\to \pi K$ and $B\to X_s \mu^+\mu^-$ do
not allow a large $Z^{\prime}$ contribution to the pure leptonic
$B_s\to\mu^+\mu^-$ decay.

\end{abstract}

\newpage
\section{Introduction}
As particle physics is entering the  era of LHC,  one  may expect  direct evidences
to  be  available to  establish whether new particles and interactions are present.
Meanwhile,  high sensitivity studies of low energy phenomena  would complement the direct
discovery physics at LHC.  The  flavor changing neutral current~(FCNC)
processes, such as $b\to s$ transitions, arise only from loop effects
within the Standard Model~(SM), and are therefore very suitable for
testing the SM and probing its various extensions. Recently, both
 CDF and D0 collaborations have announced the measurements of CP
violation parameters in $B_s$ system, with the obtained CP-violating
phase
\begin{eqnarray}\label{D0}
 \phi_s&=&-0.57^{+0.24}_{-0.30}({\rm stat})^{+0.07}_{-0.02}({\rm
 syst})\quad {\rm D0\,\,collaboration}\,\cite{D0}\,,\\
 \label{CDF}
 \phi_s&\in&[-2.82,-0.32]\,(68\% {\rm C. L.})\quad {\rm
 CDF\,\,collaboration}\,\cite{CDF}\,,
\end{eqnarray}
while within the SM this phase is expected to be
\begin{eqnarray}
 \phi_s^{SM}=-2\beta_s^{SM}=-2\arg[-V_{ts}V_{tb}^{\ast}/(V_{cs}V_{cb}^{\ast})]=-2\times (0.018\pm0.001)\,,
\end{eqnarray}
which deviates from the D0 measurement Eq.~(\ref{D0}) by more than
$2\sigma$. Combining all the available experimental information on
$\bar{B}_s-B_s$ mixing, the UTfit collaboration claims that the
divergence of $\phi_s$ between the experiment measurements and the
SM prediction is more than $3\sigma$~\cite{UTfit}.  Taking into account the deviation of  $\phi_{s}$
in a generic scenario of NP, the CKM-fitter group has found that the SM is disfavored at
$2.5\sigma$~\cite{CKMfit}. Interestingly, comparing an updated theoretical predication of   $B_{s}-\bar{B}_{s}$
 mixing with D0~\cite{D0-06} and CDF~\cite{CDF-06} early results based on $1 {fb}^{-1}$ data, the authors 
 of Ref.~\cite{Lenz} have found the mixing phase $2\sigma$ deviated from the SM expectation. Such a large observed phase, 
 if still persisting in the upcoming experimental
measurements, would indicate a signal of new physics~(NP)
manifested in $b\to s$ transitions.  In the following numerical analyses, we would use the UTfit results of $\phi_{s}$~\cite{UTfit} as benchmarks.

Motivated by the above observed anomaly, in this paper we shall
pursue possible solutions within a family non-universal $Z^{\prime}$
model~\cite{Langacker}, which could be naturally derived in certain
string constructions~\cite{string}, $E_6$ models~\cite{E6} and so
on. Searching for such an extra $Z^{\prime}$ boson is an important
mission in the experimental programs of Tevatron~\cite{Tevatron} and
LHC~\cite{LHC}. Performing constraints on the new $Z^{\prime}$
couplings through low-energy physics is, on the other hand, very
important and complementary for direct experimental searches. It is
interesting to note that, within such a specific scenario, both the
CP-violating phase problem and the well-known``$\pi K$ puzzle" in
hadronic $B\to\pi K$ decays could be resolved~\cite{Barger,Liu,
Chang}. Since both the $\bar{B}_s-B_s$ mixing and the
$B\to\pi K$ decays involve the same $b-s-Z^{\prime}$ couplings, it
is worthwhile to perform a constraint on these couplings with all
the available experimental data taken into account simultaneously.
At the same time, we could also get the allowed ranges for
flavor-conserving $u-u-Z^{\prime}$ and $d-d-Z^{\prime}$ couplings.

The FCNC $b\to s l^+l^-$~($l=e,\mu,\tau$) transition,
which gives rise to the rare inclusive $B\to X_s \mu^+ \mu^-$ and
the purely leptonic $B_s\to \mu^+ \mu^-$ decays, is another
important process to probe NP. Averaging the recent experimental
data from BABAR~\cite{BABARsuu}, Belle~\cite{Bellesuu} and
CLEO\cite{CLEOsuu}, the Heavy Flavor Averaging Group~(HFAG) presents
the following total branching ratio~\cite{HFAG}
\begin{equation}\label{BBXsmumu}
 {\cal B}(B\to X_s \mu^+\mu^-)\,=\,(4.3^{+1.3}_{-1.2})\times10^{-6}\,.
\end{equation}
As for the ones in the low ($1.0~{\rm GeV^2}<q^2<6.0~{\rm GeV^2}$)
and high ($14.4~{\rm GeV^2}<q^2<25{\rm GeV^2}$) $q^2$ regions, after
naively averaging the BABAR~\cite{BABARsuu} and
Belle~\cite{Bellesuu} measurements, we get respectively
\begin{eqnarray}
 \label{BBXsmumuL}
 &&{\cal B}^L(B\to X_s
 \mu^+\mu^-)\,=\,(1.6\pm0.5)\times10^{-6}\,,\\
 \label{BBXsmumuH}
 &&{\cal B}^H(B\to X_s \mu^+\mu^-)\,=\,(0.44\pm0.12)\times10^{-6}\,.
\end{eqnarray}
Theoretically, with the up-to-date input parameters, the SM
predictions~\cite{Buchalla:1996vs,Buras95} for the above three
observables are about $5.0\times10^{-6}$, $1.8\times10^{-6}$ and
$0.45\times10^{-6}$ respectively, which agree with the experimental
data well. It implies that such observables in
$B\to X_s \mu^+ \mu^-$, together with the measurements of
$\bar{B}_s-B_s$ mixing and hadronic $B\to\pi K$ decays, may provide
strict constraints on the new $Z^{\prime}$ couplings involving
the lepton sector.

As for the $B_s\to \mu^+\mu^-$ decay, in addition to the electro-weak
loop suppression, the decay rate is helicity suppressed in the SM
and predicted to be about $3\times
10^{-9}$~\cite{Buchalla:1996vs,Buras03,Buras09}, which is still one
order of magnitude lower than the CDF upper bound~\cite{CDFBsuu}
\begin{equation} \label{BBsmumu}
 {\cal B}(B_s\to \mu^+\mu^-)\,<\,4.7\times10^{-8}~(90\%~{\rm C.L.}).
\end{equation}
It is expected that precise measurements would  be available at the upcoming
experiments at LHC and super B factories.
As a consequence, we shall also investigate the $Z^{\prime}$
contribution to this decay mode within the parameter spaces
constrained by $\bar{B}_s-B_s$ mixing, $B\to\pi K$ and $B\to X_s
\mu^+\mu^-$ decays.

Our paper is organized as follows. In Section~2, after a brief
review of $B_q-\bar{B}_q$ mixing within the SM, we pursue possible
solutions to the $\bar{B}_s-B_s$ mixing anomaly within a family
non-universal $Z^{\prime}$ model, taking into account the constrains
from $B\to \pi K$ decays~\cite{Chang}. In Section~3, the effects of
such a NP scenario on $B\to X_s \mu^+\mu^-$ and $B_s\to \mu^+\mu^-$
decays are investigated in detail. Our conclusions are summarized in
Section~4. Appendix.~A includes all of the theoretical input
parameters.

\section{Constraints on $Z^{\prime}$ couplings from $\bar{B}_q-B_q$
mixing and $B\to \pi K$ decays}\label{Bmixing}

\subsection{Theoretical framework}
Within the SM, the effective Hamiltonian
$\mathcal{H}_{eff}^{SM}(\triangle B=2)$ for $\bar{B}_q-B_q$ mixing,
relevant for scales $\mu_b=\mathcal {O}(m_b)$ is given
by~\cite{Buchalla:1996vs}
\begin{equation}
 \mathcal{H}_{eff}^{SM}(\triangle
 B=2)=\frac{G_F^2}{16\pi^2}M_W^2(V_{tb}V_{tq}^{\ast})^2C_Q(\mu_b)Q(\triangle
 B=2)+{\rm h.c.}\,,
\end{equation}
where $Q(\triangle B=2)=(\bar{q}b)_{V-A}(\bar{q}b)_{V-A}$. Accurate
to next-to-leading order~(NLO) in QCD, the off-diagonal term
$M_{12}^{SM}(q)$ in the neutral B-meson mass matrix is given by
\begin{eqnarray}
\label{M12SM} 2m_{B_q}M_{12}^{SM}(q)&=&\langle
B_q^0|\mathcal{H}_{eff}^{SM}(\triangle B=2)|\bar{B}_q^0\rangle \nonumber\\
&=&\frac{G_F^2}{6\pi^2}M_W^2(V_{tb}V_{tq}^{\ast})^2(\hat{B}_{B_q}f_{B_q}^2)
m_{B_q}^2 \eta_{B}S_{0}(x_t)\,,
\end{eqnarray}
where $M_W$ is the mass of $W$ boson, $\hat{B}_{B_q}$ the ``bag''
parameter, and $f_{B_q}$ the B-meson decay constant. Explicit
expressions for the short-distance QCD correction function
$\eta_{B}$ and the ``Inami-Lim'' function $S_{0}(x_t)$, with
$x_t=\frac{\bar{m}_t(m_t)^2}{M_W^2}$, could be found in
Ref.~\cite{Buchalla:1996vs}.

Recently UTfit collaboration has performed a model-independent
analysis of NP effects to $\bar{B}_q-B_q$ mixing in terms of two
parameters $C_{B_q}$ and $\phi_{B_q}$, with the following
parametrization~\cite{UTfit}
\begin{equation}
\label{UTEq}
 C_{B_q}e^{2i\phi_{B_q}}\,\equiv\,\frac{\langle
 B_q|\mathcal{H}_{eff}^{full}|\bar{B}_q\rangle}
 {\langle B_q|\mathcal{H}_{eff}^{SM}|\bar{B}_q\rangle }\,
 =\,\frac{A_q^{SM}e^{i\phi_{q}^{SM}}+A_q^{NP}e^{i(2\phi_{q}^{NP}
 +\phi_{q}^{SM})}}{A_q^{SM}e^{i\phi_{q}^{SM}}}\,.
\end{equation}
Within the SM, the modulus $C_{B_q}$ and the phase $\phi_{B_q}$ are
predicted to be one and zero, respectively. Combining all the
available experimental information on $\bar{B}_q-B_q$ mixing, the
fitting results at $68\%$ and $95\%$ probabilities from
Ref.~\cite{UTfit} are listed in Table.~\ref{tab_UTresult}.
For each probability, UTfit has found two solutions for $\phi_{B_{s}}$ due to
measurement ambiguities\cite{UTfit}: one is close to, but still $3\sigma$ deviated from, the
SM expectations (denoted as S1 hereafter); another one is much more distinct from the SM and even require
dominant NP contributions (S2).
 Such large deviates may  suggest the first evidence of NP exhibited in $b\to s$ induced
processes\cite{UTfit} . So, in the following we pursue possible solutions within
a family non-universal $Z^{\prime}$ model~\cite{Langacker}.

\begin{table}[t]
 \begin{center}
 \caption{Fit results for $\bar{B}_q-B_q$ mixing parameters
 $C_{B_q}$ and $\phi_{B_q}$ by UTfit collaboration~\cite{UTfit}. The
 two solutions for $\phi_{B_s}$, S1 and S2, result from measurement
 ambiguities, see Ref.~\cite{UTfit} for details.}
 \label{tab_UTresult}
 \vspace{0.5cm}
 \doublerulesep 0.7pt \tabcolsep 0.1in
 \begin{tabular}{lccccccccccc} \hline \hline
 NP parameters & $C_{B_d}$     & $\phi_{B_d}[^{\circ}]$ & $C_{B_s}$     & $\phi_{B_s}[^{\circ}]$\,(S2 $\cup$ S1)
 \\\hline
 68\% prob.    & $0.96\pm0.23$ & $-2.9\pm1.9$           & $1.00\pm0.20$ & $(-68.0\pm4.8)\cup(-20.3\pm5.3)$ \\
 95\% prob.    & $[0.57,1.50]$ & $[-6.7,1.0]$           & $[0.68,1.51]$ & $[-77.8,-58.2]\cup[-30.5,-9.9]$ \\
  \hline \hline
 \end{tabular}
 \end{center}
 \end{table}

While the general framework for $Z^{\prime}$-induced FCNC
transitions has be formulated by Langacker and
Pl\"{u}macher~\cite{Langacker}, our discussion throughout this paper
for the $Z^{\prime}$ contributions goes under the following
simplifications: (1) neglecting kinetic mixing since it only amounts
to a redefinition of the unknown $Z^{\prime}$ couplings;
(2) neglecting the  $Z- Z^{\prime}$ mixing, which has been constrained to be tiny
 by the Z-pole measurements at LEP~\cite{DELPHI,mixAngl}, but can be
easily incorporated~\cite{Langacker,Wei}; (3) no significant
renormalization group~(RG) evolution effects between
$M_{Z^{\prime}}$ and $M_{W}$ scales; (4) although  there are
 no severe constraints on right-highted $q-{\bar  q}-Z^\prime$ couplings,
 we follow the simplification in the literature \cite{Barger, Wei, Chiang}
  and assume that right-handed couplings are flavor-diagonal and
hence real due to the hermiticity of the effective Hamiltonian,
while flavor-off- diagonal left-handed coupling terms will result in
 sizable FCNC $b_{L}-s_{L}-Z^{\prime}$ couplings.

Then, the effective Hamiltonian
$\mathcal{H}_{eff}^{Z^{\prime}}(\triangle B=2)$ induced by
$Z^{\prime}$ contribution at $M_W$ scale could be written as
\begin{equation}
 \mathcal{H}_{eff}^{Z^{\prime}}(\triangle
 B=2)=\frac{G_F}{\sqrt{2}}(B^L_{qb})^2Q(\triangle
 B=2)+{\rm h.c.}\,,
\end{equation}
where $B^L_{qb}$ is the $Z^{\prime}-b-q$ coupling, whose definition
is different from the one used in our previous paper~\cite{Chang} by
a factor $\frac{g_2M_Z}{g_1M_{Z^{\prime}}}$, with $g_1$ and $g_2$
being the gauge couplings of $Z$ and $Z^{\prime}$ bosons,
respectively. Due to our assumed simplifications, the RG running of
the Wilson coefficient induced by $Z^{\prime}$ boson is the same as
that of the SM, with the corresponding evolution matrix
$U^{LL}(\mu_b,M_W)$ given to the NLO level by~\cite{Buchalla:1996vs}
\begin{equation}
 U^{LL}(\mu_b,\mu_W)=\big{[}1+\frac{\alpha_s(\mu_b)}{4\pi}J_5\big{]}U^0
 (\mu_b,\mu_W)\big{[}1-\frac{\alpha_s(\mu_W)}{4\pi}J_5\big{]}\,,
\end{equation}
with
$U^0(\mu_b,\mu_W)=(\alpha_s(\mu_W)/\alpha_s(\mu_b))^{\frac{\gamma_Q^0}{2\beta_0}}$,
$\gamma_Q^0=4$, $\beta_0=23/3$, and $J_5=1.627$ in naive dimensional
regularization~(NDR) scheme with 5 effective quark flavors.

After some simple derivations, one can get the final $Z^{\prime}$
contribution to $M_{12}(q)$
\begin{equation}
\label{M12ZP}
 2m_{B_q}M_{12}^{Z^{\prime}}(q)=\frac{G_F}{\sqrt{2}}U_{LL}^{\prime}
 |B_{qb}^{L}|^2e^{i2\phi_q^L}\frac{8}{3}m_{B_q}^2(\hat{B}_{B_q}f_{B_q}^2)\,,
\end{equation}
with
\begin{equation}
 U_{LL}^{\prime}\equiv\big{(}\alpha_s(\mu_W)\big{)}^{\frac{\gamma^{(0)}_Q}
 {2\beta_0}}\big{[}1-\frac{\alpha_s(\mu_W)}{4\pi}J_5\big{]}.
\end{equation}
Finally, we get the total contribution to the off-diagonal mass
matrix term
\begin{equation}
M_{12}(q)=M_{12}^{SM}(q)+M_{12}^{Z^{\prime}}(q).
\end{equation}
The mass difference, which describes the strength of the
$\bar{B}_q-B_q$ mixing, is then given by $\triangle
M_q=2|M_{12}(q)|$. An early general investigation of  $Z^{\prime}$ effects in $\bar{B}_q-B_q$ mixing
could be found in Ref.\cite{Wei}.

\subsection{Numerical results and discussions}
\begin{figure}[t]
\begin{center}
\epsfxsize=15cm \centerline{\epsffile{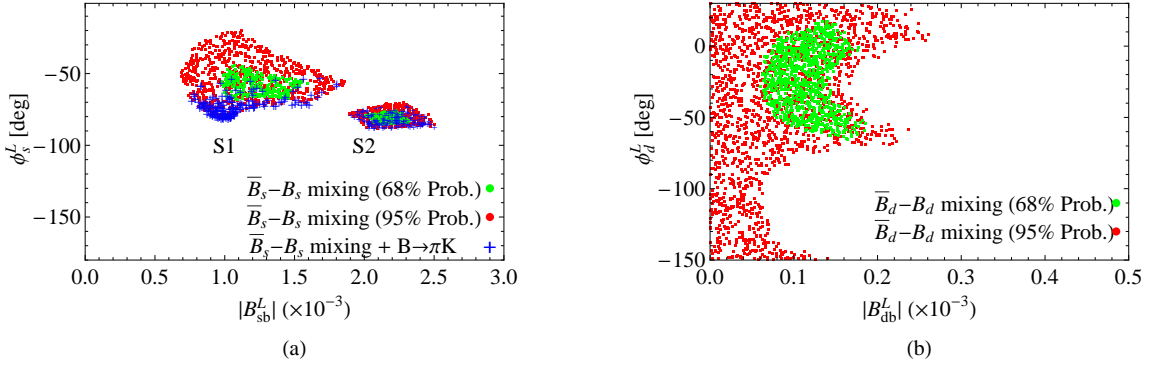}}
\centerline{\parbox{16cm}{\caption{\label{Fig1}\small The allowed
regions for the parameters $|B_{qb}^L|$ and $\phi_{q}^L$ under the
constraints from $\bar{B}_q-B_q$ mixing and $B\to\pi K$ decays. S1 and S2 correspond to the two
solutions in Table 1.}}}
\end{center}
\end{figure}

With the UTfit results at $68\%$ and $95\%$ probabilities listed
in the Table.~\ref{tab_UTresult} as constraints, respectively, we get
the allowed ranges for the $Z^{\prime}$ parameters as shown in
Fig.~\ref{Fig1}, with the corresponding numerical results given in
Table.~\ref{zprimepara1}. We find that the new $b-s-Z^{\prime}$
coupling, with a new weak phase
$\phi_s^{L}\sim-58^{\circ}$~($\phi_s^{L}\sim-80^{\circ}$) and
strength $|B_{sb}^L|$ $\sim1.2\times10^{-3}$~($2.2\times10^{-3}$)
corresponding to the UTfit result S1~(S2), is crucial to resolve the
observed $\bar{B}_s-B_s$ mixing phase anomaly.
 On the other hand, the $\Delta M_{d}$  is well measured and in good agreement with the SM predictions,   
 the strength of $b-d-Z^{\prime}$ coupling involved in  $B^{0}_{d}-\bar{B^{0}_{d}}$ mixing  should be much
  weaker than the one of the SM box diagrams. 
 Numerically, $|B^{L}_{db}|$ is found to to be about  $1.2\times10^{-4}$. 
 Combining with the constraints by $B^{0}_{s}-\bar{B^{0}_{s}}$  mixing and $B\to \pi K$ decays,   
 we  find that the relative strength, $|B_{db}^{L}/B_{sb}^{L}| \sim \mathcal{O}(10^{-1})$, 
  is quite similar to the hierarchy of CKM matrix elements within the SM, $|V_{td}^{\ast}V_{tb}/V_{ts}^{\ast}V_{tb}|\sim0.2$.
On the theoretical side, it is noted that such a hierarchy is not
required by the $Z^{\prime}$ model itself.  Although the $Z^{\prime}$ model
 considered here is not a model of  Minimal Flavor Violation (MFV)  type,   the low energy effective Hamiltonian resulting from
the  flavor-changing $Z^{\prime}$ couplings  also obey  the so-called MFV
hypothesis~\cite{D'Ambrosio:2002ex,Buras:2003jf} driven by the experimental data available so far,
which may imply that the flavor-changing interactions in the $Z^{\prime}$ model are also linked 
to the known structure of the SM Yukawa couplings.

\begin{table}[t]
 \begin{center}
 \caption{Numerical results for the parameters $|B_{qb}^L|$ and
 $\phi_{q}^L$ under the constraints from $\bar{B}_q-B_q$ mixing
 only. For each solution, the first~(second) row corresponds to the
 constraints from $C_{B_q}$ and $\phi_{B_q}$ at $68\%$~($95\%$) probability. }
 \label{zprimepara1}
 \vspace{0.5cm}
 \doublerulesep 0.7pt \tabcolsep 0.1in
 \begin{tabular}{lccccccccccc} \hline \hline
 Solutions               & $|B_{sb}^L|(\times10^{-3})$& $\phi_{s}^L[^{\circ}]$ & $|B_{db}^L|(\times10^{-3})$ & $\phi_{d}^L[^{\circ}]$\\\hline
 S1                      & $1.24\pm0.16$              & $-58\pm6$              & $0.12\pm0.03$               & $-23\pm21$              \\
                         & $1.18\pm0.29$              & $-52\pm13$             &$\leqslant0.26$             & arbitrary             \\\hline
 S2                      & $2.17\pm0.07$              & $-80\pm2$              &---&  ---&     \\
                         & $2.19\pm0.14$              & $-80\pm4$              &---&  ---&    \\
  \hline \hline
 \end{tabular}
 \end{center}
 \end{table}

From Eqs.~(\ref{M12SM}), (\ref{UTEq}) and (\ref{M12ZP}), we find
that the parameters $C_{B_q}$ and $\phi_{B_q}$ are independent of
the theoretical uncertainties associated with the non-perturbative
factor $\hat{B}_{B_q}f_{B_q}^2$ within such a family non-universal
$Z^{\prime}$ model under our assumed simplifications. However, to
get the mass difference $\triangle M_q$, such uncertainties are
unavoidable. With the relevant input parameters listed in the
Appendix~A, the final numerical results for $\triangle M_q$ are
listed in Table~\ref{MassDiff}. It can be seen that, after including
the $Z^{\prime}$ contributions, our predictions for $\triangle M_q$
also agree with the experiment data, taking into account the
respective theoretical uncertainties.

\begin{table}[t]
 \begin{center}
 \caption{Numerical results for the mass difference $\triangle
 M_q$~(${\rm ps}^{-1}$).}
 \label{MassDiff}
 \vspace{0.5cm}
 \doublerulesep 0.7pt \tabcolsep 0.1in
 \begin{tabular}{lccccccccccc} \hline \hline
 Solutions               & Exp.~\cite{HFAG} & SM             &S1             & S2\\\hline
 $\triangle M_d$         & $0.508\pm0.005$ & $0.525\pm0.057$&$0.522\pm0.077$& ---             \\
 $\triangle M_s$         & $17.77\pm0.12$  & $18.18\pm1.47$ &$17.14\pm2.30$ &$17.42\pm2.40$     \\
  \hline \hline
 \end{tabular}
 \end{center}
 \end{table}

In our pervious paper~\cite{Chang}, we found that a nontrivial new
weak phase $\phi_{s}^{L}\sim -86^{\circ}$ associated with the
$b-s-Z^{\prime}$ coupling is helpful to resolve the so-called ``$\pi
K$ puzzle'', which is similar to our present fitting result
$\phi_s^{L}\sim-58^{\circ}$~in S1~($\phi_s^{L}\sim-80^{\circ}$~in
S2) from $\bar{B}_s-B_s$ mixing. However, as found in
Ref.~\cite{Chang}, the range $\phi_s^{L}>-50^{\circ}$ is almost
excluded by the CP-averaged branching ratios and direct CP
asymmetries of $B\to\pi K$ decays. So, it is very necessary and
interesting to re-evaluate the ranges of $Z^{\prime}$ couplings
under the constraints from $\bar{B}_s-B_s$ mixing and $B\to\pi K$
decays simultaneously.

Like the Case~IV in Ref.~\cite{Chang}, we give up any
simplifications on the flavor-diagonal $u-u-Z^{\prime}$ and
$d-d-Z^{\prime}$ couplings, and use the QCD
factorization~(QCDF)~\cite{Beneke1} approach to calculate the
amplitudes of $B\to\pi K, \pi K^{\ast}$ and $\rho K$ decays. As for
the end-point divergence appearing in twist-3 spectator and
annihilation amplitudes, instead of the parametrization scheme, we
quote an infrared finite dynamical gluon propagator derived by
Cornwall~\cite{Cornwall} to regulate it. Explicitly we quote
$m_g=0.50\pm0.05\,{\rm GeV}$, which is a reasonable choice so that
most of the observables for $B\to\pi K, \pi K^{\ast}$ and $\rho K$
decays are in good agreement with the experimental
data~\cite{YDYang}. In this way, we find that the time-like
annihilation amplitude could contribute a large strong-interaction
phase, while the space-like spectator-scattering amplitude is
real~\cite{YDYang}. The explicit comparison for the two schemes have
been systemically discussed in our previous
papers~\cite{Chang,YDYang}. Although numerically these two schemes
 have some differences, both of their predictions are consistent with
most of the experimental data within errors.

\begin{figure}[t]
\begin{center}
\epsfxsize=15cm \centerline{\epsffile{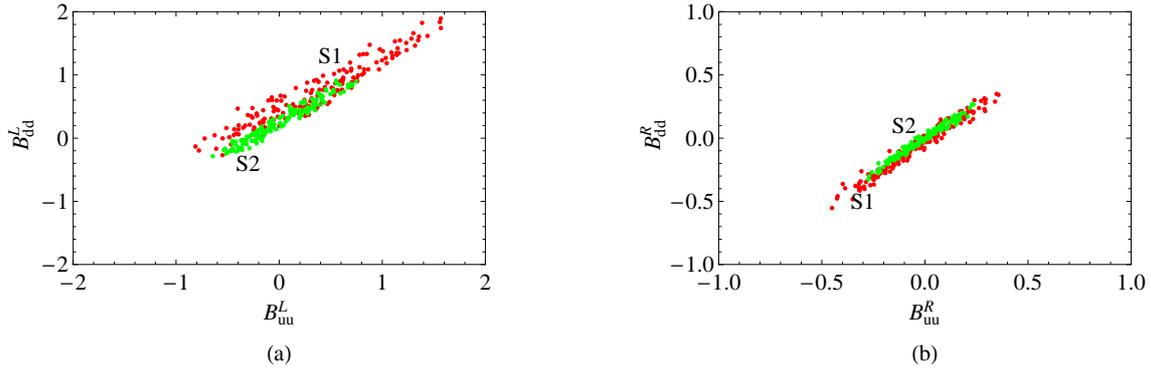}}
\centerline{\parbox{16cm}{\caption{\label{Fig2}\small The allowed
regions for the parameters $B_{uu,dd}^{L,R}$ under the constraints
from $C_{B_s}$, $\phi_{B_s}$~($95\%$ prob. only) and $B\to\pi K$
decays.}}}
\end{center}
\end{figure}

\begin{table}[t]
 \begin{center}
 \caption{Numerical results for the parameters $|B_{sb}^L|$,
 $B_{uu,dd}^{L,R}$ and $\phi_{s}^L$ under the constraints from
 $\bar{B}_s-B_s$ mixing and $B\to\pi K$ decays. The other captions
 are the same as the ones in Table~\ref{zprimepara1}.}
 \label{zprimepara2}
 \vspace{0.5cm}
 \small
 \doublerulesep 0.7pt \tabcolsep 0.05in
 \begin{tabular}{lccccccccccc} \hline \hline
 Solutions               & $|B_{sb}^L|(\times10^{-3})$ & $\phi_{s}^L[^{\circ}]$ &$B_{uu}^L$    &$B_{uu}^R$     & $B_{dd}^L$   & $B_{dd}^R$ \\\hline
 S1                      &$1.18\pm0.16$                &$-62\pm5$               &$0.66\pm0.38$ &$-0.13\pm0.12$ &$0.88\pm0.36$ &$-0.18\pm0.14$\\
                         &$1.09\pm0.22$                &$-72\pm7$               &$0.34\pm0.55$ &$-0.04\pm0.18$ &$0.70\pm0.48$ &$-0.07\pm0.20$\\\hline
 S2                      &$2.19\pm0.06$                &$-81\pm2$               &$-0.02\pm0.34$ &$0.01\pm0.12$ &$0.22\pm0.32$ &$0.01\pm0.12$     \\
                         &$2.20\pm0.15$                &$-82\pm4$               &$0.02\pm0.34$ &$-0.01\pm0.12$ &$0.27\pm0.32$ &$-0.04\pm0.24$ \\
  \hline \hline
 \end{tabular}
 \end{center}
 \end{table}

Including the constraints from $\bar{B}_s-B_s$ mixing and $B\to\pi
K$ decays, the final allowed ranges for the $Z^{\prime}$ couplings
are presented  in Figs.~\ref{Fig1} and \ref{Fig2}. As shown in
Fig.~\ref{Fig1}, the range of $\phi_s^{L}$ in S1 is now further
restricted with the constraints from $B\to\pi K$ decays
included~({\it i.e.}, the range $\phi_s^{L}>-50^{\circ}$ is now
excluded), while their effect for  S2 case is tiny. Our numerical
results for $Z^{\prime}$ couplings are summarized  in
Table.~\ref{zprimepara2}.

 For evaluating $B\to \pi K$ decays, since the $Z^{\prime}$ mediated effects can occur not only in the coefficients of the electro-weak 
 penguin operators but also in the strong penguin ones, we do not assume  $B_{dd}^R/B_{uu}^{R}=e_{d}/e_{u}=-1/2$ 
 (if assumed, the $Z^{\prime}$ effects in terms of  the Wilson coefficients of the SM QCD penguin operators will then vanish, 
as usually adopted in the literature, see for example Refs.~\cite{Barger,Liu,Chang}). So, compared
with $|B_{dd}^R|=|-B_{uu}^{R}/2|<0.1$~\cite{Liu}, our fitting result
in Fig.~\ref{Fig2} shows that a larger ranges for
$B_{uu,dd}^{L,R}$ are still allowed. Furthermore, as can be seen
from Fig.~\ref{Fig2} and Table.~\ref{zprimepara2},
$B_{dd}^{L}$~($B_{uu}^{R}$) is a bit larger than
$B_{uu}^{L}$~($B_{dd}^{R}$). However, remembering that the
$Z^{\prime}$ contribution to the electro-weak penguin coefficient
$\Delta C_7$ is the combination $B_{uu}^R-B_{dd}^R$, our numerical
result always satisfies $|B_{uu}^R-B_{dd}^R|<0.13$, which is still
consistent with the result $3|B_{dd}^R|<0.3$ obtained very recently by Barger et al.~\cite{Liu}.

\section{Constraints on $Z^{\prime}$ couplings from  $B_s\to\mu^+\mu^-$ and $B\to X_s
\mu^+\mu^-$}\label{Brare}

\subsection{Theoretical Framework}\label{B2Xsll}
Within the SM, after dropping the negligible charm contributions,
the effective Hamiltonian for purely leptonic $B_s\to l^+ l^-$ decay
is given as~\cite{Buchalla:1996vs,BuchallaB2ll,Buras94}
\begin{equation}\label{SMHbsll}
 {\cal H}_{eff}^{SM}(B_s\to
 l^+l^-)=-\frac{G_F}{\sqrt{2}}\frac{\alpha}{2\pi\sin^2\theta_{W}}
 V_{tb}V^{\ast}_{ts}Y(x_t)(\bar{s}b)_{V-A}(\bar{l}l)_{V-A}+{\rm h.c.}\,,
\end{equation}
where $\alpha=\frac{e^2}{4\pi^2}=1/137$,
$\sin^2\theta_{W}=0.23119$~\cite{PDG08}, and the function $Y(x_t)$
is defined as \cite{Buchalla:1996vs,BuchallaB2ll}
\begin{eqnarray}
   Y(x_t)&=&Y_0(x_t)+\frac{\alpha_s}{4\pi}Y_1(x_t)\,,\nonumber\\[0.2cm]
 Y_0(x_t)&=&\frac{x_t}{8}\big[\frac{x_t-4}{x_t-1}+\frac{3x_t}{(x_t-1)^2}\ln{x_t}\big]\,,\\[0.2cm]
 Y_1(x_t)&=&\frac{4x_t+16x_t^2+4x_t^3}{3(1-x_t)^2}+\frac{2x_t+x_t^3}{(1-x_t)^2}\rm{Li_2}
           (1-x_t)+8x_t\frac{\partial Y_0(x_t)}{\partial x_t}\ln\frac{\mu_t^2}{M_W^2}\nonumber\\
         & &-\frac{4x_t-10x_t^2-x_t^3-x_t^4}{(1-x_t)^3}\ln{x_t}+\frac{2x_t-14x_t^2+x_t^3-x_t^4}
         {2(1-x_t)^3}\ln^2{x_t}\,.\nonumber
\end{eqnarray}

Within our approximations for the non-universal $Z^{\prime}$
couplings, the effective Hamiltonian for $b\to sl^+l^-$ transition
induced by the new $Z^{\prime}$ boson could be written as
\begin{equation}\label{ZPHbsll}
 {\cal H}_{eff}^{Z^{\prime}}(b\to sl^+l^-)=-\frac{2G_F}{\sqrt{2}}
 V_{tb}V^{\ast}_{ts}\Big[-\frac{B_{sb}^{L}B_{ll}^{L}}{V_{tb}V^{\ast}_{ts}}
 (\bar{s}b)_{V-A}(\bar{l}l)_{V-A}-\frac{B_{sb}^{L}B_{ll}^{R}}{V_{tb}V^{\ast}_{ts}}
 (\bar{s}b)_{V-A}(\bar{l}l)_{V+A}\Big]+{\rm h.c.}\,.
\end{equation}
Then, the full expression for the branching ratio of $B_s\to l^+l^-$
is
\begin{eqnarray}\label{BrBsmumu}
 {\cal B}(B_s\to l^+l^-)&=&\tau_{B_s}\frac{G_F^2}{4\pi}f_{B_s}^2m_l^2m_{B_s}
 \sqrt{1-\frac{4m_l^2}{m_{B_s}^2}}|V_{tb}V^{\ast}_{ts}|^2\nonumber\\
                        & &\times{\Big|}\frac{\alpha}{2\pi \sin^2\theta_W}Y(x_t)
 -2\frac{B_{sb}^{L}(B_{ll}^L-B_{ll}^R)}{V_{tb}V^{\ast}_{ts}}{\Big|}^2\,.
\end{eqnarray}
To clarify the differences between Eq.~(\ref{BrBsmumu}) and the ones  in the literature\footnote{ A sign  in Eq.~(B1) in Ref.~\cite{Liu} and
Eq.~(15) in Ref.~\cite{Chiang} mistyped,  and the interference terms missed in Eq.~(B1) of  the first version Ref.~\cite{Liu}.  We thank T. Liu and C.W. Chiang for confirmation},  a detailed derivation of Eq.~(\ref{BrBsmumu}) is presented in Appendix~B.


The SM effective Hamiltonian for rare $b\to sl^+ l^-$ decay at scale
$\mu$ is given by
\begin{equation}
 {\cal H}_{eff}^{SM}(b\to sl^+l^-)=-\frac{G_F}{\sqrt{2}}V_{tb}V^{\ast}_{ts}
 \Big[\sum_{i=1}^{8}C_i(\mu)Q_i+C_{9V}(\mu)Q_{9V}+C_{10A}(\mu)Q_{10A}\Big]+{\rm h.c.}\,.
\end{equation}
Here we choose the operator basis given by
Refs.~\cite{Buchalla:1996vs,Buras95}, in which
\begin{equation}
 Q_{9V}=(\bar{s}b)_{V-A}(\bar{l}l)_V\,,\quad Q_{10A}=(\bar{s}b)_{V-A}(\bar{l}l)_A\,.
\end{equation}
Modifying the $Z^{\prime}$-induced effective Hamiltonian
Eq.~(\ref{ZPHbsll}) to the above form, we find that the $Z^{\prime}$
effects can be represented as some modifications of the Wilson
coefficient of the corresponding operators. To this end, the initial
conditions for the coefficients $C_{9V}$ and $C_{10A}$ at the
matching scale $\mu=M_W$ are given as
\begin{eqnarray}
 C_{9V,10A}(M_W)&=&C_{9V,10A}^{SM}(M_W)+{\triangle}C_{9V,10A}^{\prime}(M_W)\,,\\
 {\triangle}C_{9V}^{Z^{\prime}}(M_W)&=&-2\frac{B_{sb}^L}{V_{tb}V^{\ast}_{ts}}(B_{ll}^{L}+B_{ll}^{R})\,,\\
 {\triangle}C_{10A}^{Z^{\prime}}(M_W)&=&2\frac{B_{sb}^L}{V_{tb}V^{\ast}_{ts}}(B_{ll}^{L}-B_{ll}^{R})\,,
\end{eqnarray}
with $C_{9V,10A}^{SM}$ and ${\triangle}C_{9V,10A}^{\prime}(M_W)$
denoting the SM and NP parts respectively. The RG running of these
Wilson coefficients has been detailed in
Refs.~\cite{Buchalla:1996vs,Buras95,Buras94}.

Introducing the normalized dilepton invariant mass
$\hat{s}=(p_{l^+}+p_{l^-})^2/m_b^2$, the differential decay rate
with respect to $\hat{s}$ for $b\to sl^+ l^-$ reads
\begin{equation}
 R(\hat{s})\equiv\frac{\frac{d}{d\hat{s}}{\cal B}(b\to s l^+l^-)}{{\cal B}(b\to c l^-\bar{\nu})}
           =\frac{\alpha^2}{4\pi^2}\frac{|V_{ts}^{\ast}V_{tb}|^2}{|V_{cb}|^2}\frac{(1-\hat{s})^2}
           {f(\chi)\kappa(\chi)}\sqrt{1-\frac{4t^2}{\hat{s}}}D(\hat{s})\,,
\end{equation}
with
\begin{eqnarray}
D(\hat{s})&=&(1+2\hat{s})(1+\frac{2t^2}{\hat{s}})|\widetilde{C}_9^{eff}|^2\,
+\,4(1+\frac{2}{\hat{s}})(1+\frac{2t^2}{\hat{s}})|{C}_7^{eff}|^2\nonumber\\
&&+\,\big[(1+2\hat{s})+\frac{2t^2}{\hat{s}}(1-4\hat{s})\big]|
\widetilde{C}_{10}|^2+12(1+\frac{2t^2}{\hat{s}}){C}_7^{eff} {\rm
Re}(\widetilde{C}_9^{eff\ast})\,,
\end{eqnarray}
where $t=m_l/m_b$, $\chi=m_c/m_b$ and ${\cal B}(B\to X_c l^-
\nu_l)=(10.1\pm0.4)\%$~\cite{PDG08}. The phase-space factor
$f(\chi)$ and the 1-loop QCD correction factor $\kappa(\chi)$ for
$B\to X_c l^-\bar{\nu}$ decay are given respectively
by~\cite{factorfk}
\begin{eqnarray}
f(\chi)&=&1-8\chi^2+8\chi^6-\chi^8-24\chi^4\ln\chi\,,\\
\kappa(\chi)&=&1-\frac{2\alpha_s(\mu)}{3\pi}\Big[\big(\pi^2-\frac{31}{4}\big)
(1-\chi)^2+\frac{3}{2}\Big]\,.
\end{eqnarray}
The effective coefficient $\widetilde{C}_9^{eff}$ is defined as \cite{Buras95}
\begin{eqnarray}\label{hsq}
\widetilde{C}_9^{eff}&=&\widetilde{C}_9\eta(\hat{s})+  h(\chi,\hat{s})(3C_1+C_2+3C_3+C_4+3C_5+C_6)-\frac{1}{2}h(1,\hat{s})
 (4C_4+4C_4+3C_5+C_6)\nonumber\\
            &&-\frac{1}{2}h(0,\hat{s})(C_3+3C_4)+\frac{2}{9}(3C_3+C_4+3C_5+C_6)\,
  \end{eqnarray}
 where  the function  $\eta(\hat{s})$ in the first term represents one gluon corrections to the matrix element of  $Q_{9V}$, 
 the other  terms arise from the insertions of four-quark operators (indicated by the $C_{i}$) to the one-loop matrix element 
 of  $Q_{9V}$  \cite{Buras95,Misiak,Kuhn}  
\begin{eqnarray}
 \eta(\hat{s})&=&1+\frac{\alpha_s(\mu)}{\pi}\big[-\frac{2}{9}\pi^2-\frac{4}{3}
 \text{Li}_2(\hat{s})-\frac{2}{3}\ln\hat{s}\ln(1-\hat{s})-\frac{5+4\hat{s}}
 {3(1+2\hat{s})}\ln(1-\hat{s})\nonumber\\
&&-\frac{2\hat{s}(1+\hat{s})(1-2\hat{s})}{3(1-\hat{s})^2(1+2\hat{s})}\ln\hat{s}
+\frac{5+9\hat{s}-6\hat{s}^2}{6(1-\hat{s})(1+2\hat{s})}\big]\,,\\[0.3cm]
 h(\chi,\hat{s})&=&-\frac{8}{9}\ln\frac{m_b}{\mu}-\frac{8}{9}\ln{\chi}+\frac{8}{27}
 +\frac{4}{9}x\nonumber\\
&&-\frac{2}{9}(2+x)|1-x|^{\frac{1}{2}}\left\{
\begin{array}{ll}
\ln|\frac{\sqrt{1-x}+1}{\sqrt{1-x}-1}|-i\pi\quad  &\text{for}\,x\equiv\frac{4\chi^2}
{\hat{s}}<1\,,\\
2\arctan\frac{1}{\sqrt{x-1}}\quad  &\text{for}\,x\equiv\frac{4\chi^2}{\hat{s}}>1\,,
\end{array}
\right.\\[0.3cm]
 h(0,\hat{s})&=&\frac{8}{27}-\frac{8}{9}\ln\frac{m_b}{\mu}-\frac{4}{9}\ln\hat{s}
 +\frac{4}{9}i\pi\,.
\end{eqnarray}
Besides these well defined short distance contributions, there are long distance corrections related to 
the  $c\bar c$ intermediate states. Phenomenologically they are estimated with Briet-Wigner approximation\cite{Ali91,Wise96,Ali97}  
which results in a modification to $\widetilde{C}_{9}^{eff}$ by 
\begin{eqnarray} 
Y_{res}(\hat{s})=\frac{3\pi}{\alpha^2}\kappa(3C_1+C_2+3C_3+C_4+3C_5+C_6)
 \sum_{V_n=\Psi(nS)}\frac{\Gamma(V_n\to l^+l^-)m_{V_n}}
 {m_{V_n}^2-\hat{s}m_b^2-im_{V_n}\Gamma_{V_n}}.
\end{eqnarray}
Usually the  factor $\kappa\simeq 2.3$ (more precisely 2.3 times an arbitrary strong phase\footnote{we thank the 
referee  for bring this point to us}\cite{Wise96} ) is introduced phenomenologically  to include the known factorizable and the unknown
nonfactorizable contributions to account for 
the present experimental data on $B\to \Psi(nS) X_{s}$ decays.
This approximation, however, will cause a double counting of partonic and hadronic  degrees of freedom\cite{KS96}. 
To avoid the double counting,  it has been suggested that the long-distance effects 
could be estimated by means of experimental data on $R_{c}(s)=\sigma(e^{-}e^{+}\to c{\bar c})/\sigma(e^{-}e^{+} \to \mu^{+}\mu^{-})$ using a dispersion relation~\cite{KS96}(KS approach).  In KS approach, only factorizable effect $i.e.$ the $c\bar{c}$ in color singlet, could be estimated with $R_{c}(s)$.   It still needs the phenomenological enhancement factor $\kappa$ to model possible nonfactorizable effects
 to match  the aforementioned large rate of $B\to\Psi(\Psi') X_{s}$. However, as discussed in detail in Ref. \cite{BBNSEJPC}, 
 unlike the $c\bar{c}$ contribution to the  $e^{-}e^{+}\to$ hadrons cross section where  the imaginary part of a current-current 
 correlator is integrated over phase space,  the huge charm-resonance contributions to $B\to X_{s} \ell^{+}\ell^{-}$ are related to a drastic 
failure of quark-hadron duality in the narrow-resonance region for integrating the absolute  {\it square} of  the correlator  over the 
phase space.  Further detailed discussions on these non-perturbative effects could be found in Ref.~\cite{BBNSEJPC}. In this paper, we concentrate on the short-distance effects and use the low- and the high-$s$ data, i.e., 
 away from the $\Psi$ and $\Psi'$ peaks, to constrain NP effects, and ignore the  resonance effects. It should be noted that, unlike $\Psi$ and $\Psi^{\prime}$(large data samples, known structures, etc.),  the backgrounds due to higher  $J^{PC}=1^{--}$ charmonium  resonances in the high-$s$ region may be very hard to be vetoed experimentally.  Although their effects are expected to be  much smaller than 
 the former ones, they still cause sizable uncertainties which are  hard to be estimated \cite{Tobias}. In a recent study of exclusive 
 $B\to K \ell^{+}\ell^{-}$  decay\cite{bucha}, it is argued that duality violations from the higher resonances in high- $s$ region  are 
 at a moderate level which may spoil the  precision of theoretical predictions for (partially) integrated branching ratios of  $B\to K \ell^{+}\ell^{-}$ at the level of several percentage.   In view of the large uncertainties included in our numerical analyses, we may expect the effects of higher $c\bar c$  resonances would not alter our conclusion much.


Finally, the normalized forward-backward~(FB) asymmetry distribution is
defined as
\begin{eqnarray}
A_{FB}(\hat{s})=\frac{\int_0^1dz\frac{d^2\Gamma}{d\hat{s}dz}
-\int_{-1}^0dz\frac{d^2\Gamma}{d\hat{s}dz}}{\int_0^1dz\frac{d^2\Gamma}
{d\hat{s}dz}+\int_{-1}^0dz\frac{d^2\Gamma}{d\hat{s}dz}}
=-3\sqrt{1-\frac{4t^2}{\hat{s}}}\frac{E(\hat{s})}{D(\hat{s})},
\end{eqnarray}
with
\begin{eqnarray}
E(\hat{s})={\rm
Re}(\widetilde{C}_9^{eff}\widetilde{C}_{10}^{\ast}\hat{s}+2{C}_7^{eff}\widetilde{C}_{10}^{\ast}).
\end{eqnarray}

\subsection{Numerical analyses and discussions}
With the relevant theoretical formulas collected in
section~\ref{B2Xsll} and the input parameters summarized in the
Appendix~A, we now proceed to present our numerical analyses and
discussions. The rare $B\to X_s \mu^+\mu^-$ and $B_s\to\mu^+\mu^-$
decays involve not only the coupling $B_{sb}^{L}$, which has been
severely constrained by $\bar{B}_s-B_s$ mixing and $B\to\pi K$
decays discussed in section~\ref{Bmixing}, but also the unrestricted
$\mu-\mu-Z^{\prime}$ couplings, $B_{\mu\mu}^{L}$ and
$B_{\mu\mu}^{R}$. So, our analyses are further divided into the
following three cases with different simplifications for our
attention, namely
\begin{itemize}
\item Case I: with $B^{L}_{\mu\mu}$ arbitrary, while taking $B_{\mu\mu}^{R}=0$;

\item Case II: with $B^{R}_{\mu\mu}$ arbitrary, while taking $B_{\mu\mu}^{L}=0$;

\item Case III: with both $B^{L}_{\mu\mu}$ and $B^{R}_{\mu\mu}$
arbitrary.
\end{itemize}

In the following discussions, we quote the fitting results for
$|B_{sb}^{L}|$ and $\phi_s^{L}$ under the constraints from
$C_{B_q}$, $\phi_{B_s}$~(95\% prob.) and $B\to\pi K$ decays as
inputs. In our numerical evaluations, we don't consider the LD
contribution to $\widetilde{C}_9^{eff}$. In each case, our fitting
is performed with the experimental data on ${\cal B}(B\to X_s
\mu^+\mu^-)$, ${\cal B}^{H}(B\to X_s \mu^+\mu^-)$ and ${\cal
B}^{L}(B\to X_s \mu^+\mu^-)$ varying randomly within their
respective $1\sigma$ error bars, while the theoretical uncertainties
are obtained by varying the input parameters within the regions
specified in Appendix~A. Moreover, we leave ${\cal
B}(B_s\to\mu^+\mu^-)$, $A_{FB}(B\to X_s \mu^+\mu^-)$ and
$A_{FB}^{L,H}(B\to X_s \mu^+\mu^-)$ as our theoretical prediction,
which could be tested by more precise measurements in the coming
years.

\begin{figure}[t]
\begin{center}
\epsfxsize=15cm \centerline{\epsffile{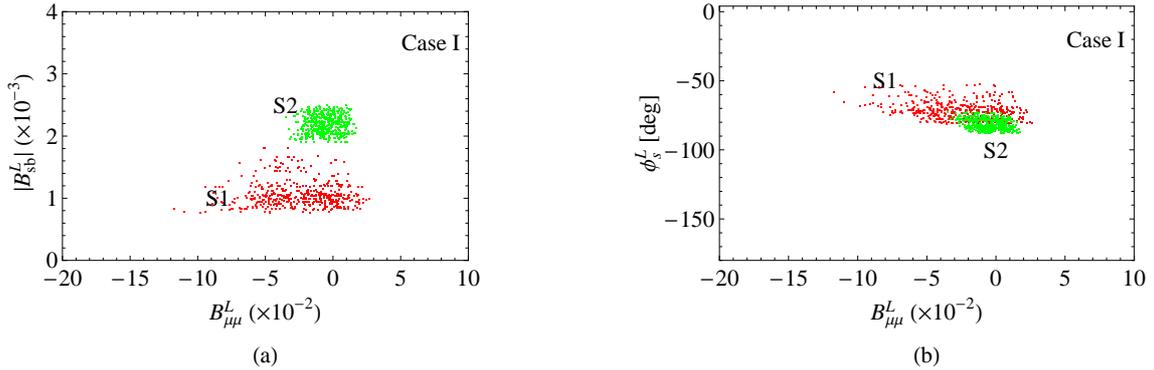}}
\centerline{\parbox{16cm}{\caption{\label{CaseI}\small The allowed
regions for the parameters $B_{\mu\mu}^{L}$ in Case~I.}}}
\end{center}
\end{figure}

\begin{table}[t]
 \begin{center}
 \caption{Numerical results for the parameters $B_{\mu\mu}^{L}$ and
 $B_{\mu\mu}^{R}$~(in unit of $\times10^{-2}$).}
 \label{zprimepara3}
 \vspace{0.5cm}
 \small
 \doublerulesep 0.7pt \tabcolsep 0.05in
 \begin{tabular}{lccccccccccc} \hline \hline
 Cases            &\multicolumn{2}{c}{Case I}  & \multicolumn{2}{c}{Case II} &\multicolumn{2}{c}{Case III} \\
                  & S1          & S2           & S1           &S2            & S1            & S2              \\\hline
 $B_{\mu\mu}^{L}$ &$-2.5\pm2.7$ &$-0.55\pm1.0$  &---           &---           &$-2.7\pm2.5$ &$-0.59\pm0.93$\\
 $B_{\mu\mu}^{R}$ &---          &---           &$0.78\pm2.0$ &$0.23\pm0.97$  &$0.61\pm2.4$  &$0.19\pm0.88$\\
 \hline \hline
 \end{tabular}
 \end{center}
 \end{table}

\begin{figure}[t]
\begin{center}
\epsfxsize=15cm \centerline{\epsffile{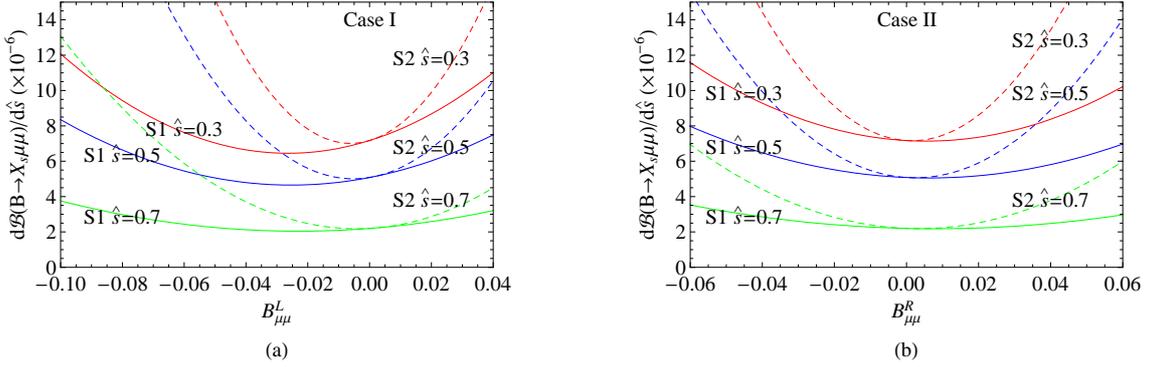}}
\centerline{\parbox{16cm}{\caption{\label{FigDBBuuLR}\small The
dependence of $d{\cal B}(B\to X_s \mu^+\mu^-)/d\hat{s}$ on
$B_{\mu\mu}^{L(R)}$ at $\hat{s}=0.3$, $0.5$ and $0.7$ with
$|B_{sb}^{L}|=1.09\times10^{-3}$($2.20\times10^{-3}$) and
$\phi_s^L=-72^{\circ}$($-82^{\circ}$) in Case~I S1~(S2) and Case~II
S1~(S2). }}}
\end{center}
\end{figure}

\begin{figure}[t]
\begin{center}
\epsfxsize=15cm \centerline{\epsffile{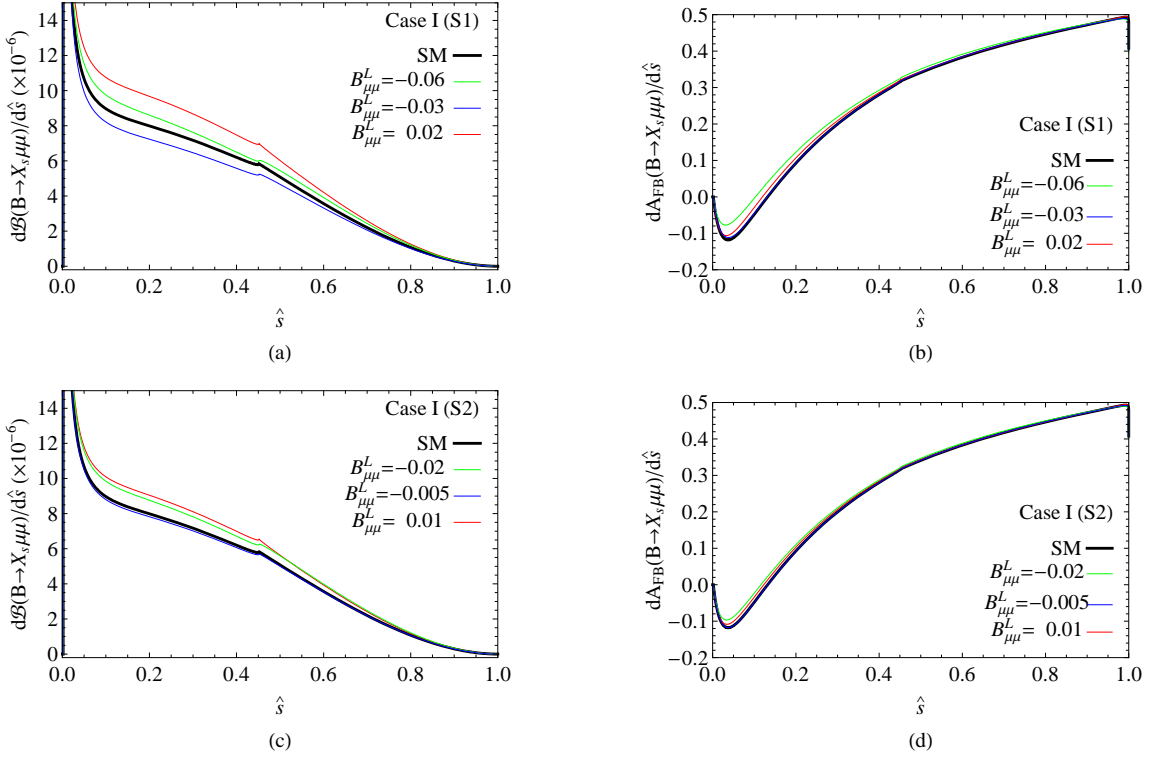}}
\centerline{\parbox{15cm}{\caption{\label{FigDBAbfSCaseI}\small The
dependence of $d{\cal B}(B\to X_s \mu^+\mu^-)/d\hat{s}$ and
$dA_{FB}(B\to X_s \mu^+\mu^-)/d\hat{s}$ on $\hat{s}$ with
$|B_{sb}^{L}|=1.09\times10^{-3}$($2.20\times10^{-3}$),
$\phi_s^L=-72^{\circ}$($-82^{\circ}$), and $B_{\mu\mu}^{R}=0$.}}}
\end{center}
\end{figure}

\subsubsection*{Case~I: with $B^{L}_{\mu\mu}$ arbitrary, while taking $B_{\mu\mu}^{R}=0$.}
In order to investigate the effects of $B^{L}_{\mu\mu}$, we neglect
the $Z^{\prime}$ contributions involving $B_{\mu\mu}^{R}$ in this
case. Corresponding to the two solutions S1 and S2 for $B_{sb}^{L}$,
we obtain two allowed regions for $B^{L}_{\mu\mu}$ as shown in
Fig.~\ref{CaseI}, and the corresponding numerical results are listed
in Table~\ref{zprimepara3}. Our predictions for ${\cal B}(B\to X_s
\mu^+\mu^-)$, $A_{FB}(B\to X_s \mu^+\mu^-)$, including the results
at both  low and high $q^2$ regions, and ${\cal
B}(B_s\to\mu^+\mu^-)$ are given in Table~\ref{predictions}. Due to
the fact that the SM predictions, ${\cal B}(B\to
X_s\mu^+\mu^-)=(5.0\pm0.3)\times10^{-6}$, ${\cal B}^{L}(B\to
X_s\mu^+\mu^-)=(1.8\pm0.1)\times10^{-6}$, ${\cal B}^{H}(B\to
X_s\mu^+\mu^-)=(0.45\pm0.06)\times10^{-6}$, and ${\cal
B}(B_s\to\mu^+\mu^-)=(3.1\pm0.2)\times10^{-9}$, agree quite well
with the experimental measurements as given by
Eqs.~(\ref{BBXsmumu}), (\ref{BBXsmumuL}), (\ref{BBXsmumuH}), and
(\ref{BBsmumu}), the parameter space with $B^{L}_{\mu\mu}\sim0$ is
still allowed. However, as shown latter, nonzero
$B^{L}_{\mu\mu}$ may have significant  impacts on the other
observables.

With the central values of theoretical input parameters,
Fig.~\ref{FigDBBuuLR}~(a) shows the dependence of $d{\cal B}(B\to
X_s \mu^+\mu^-)/d\hat{s}$ on $B_{\mu\mu}^{L}$ at different
$\hat{s}$, from which we can see that the minimal value of the
differential decay rate appears at
$B_{\mu\mu}^{L}\sim-0.03$~($-0.005$) in S1~(S2). Furthermore, the
dependence of $d{\cal B}(B\to X_s \mu^+\mu^-)/d\hat{s}$ and
$dA_{FB}(B\to X_s \mu^+\mu^-)/d\hat{s}$ on $\hat{s}$, with different
values of $B_{\mu\mu}^{L}$, is shown in Fig.~\ref{FigDBAbfSCaseI}.
From Fig.~\ref{FigDBAbfSCaseI}~(a), one can find that
$d{\cal B}(B\to X_s \mu^+\mu^-)/d\hat{s}$ is reduced at
$B_{\mu\mu}^{L}\sim-0.03$ but enhanced at $B_{\mu\mu}^{L}\sim-0.06$
and $0.02$ in S1. In S2, with $B_{\mu\mu}^{L}=-0.02$, the
$Z^{\prime}$ contributions induced by $B_{\mu\mu}^{L}$ could enhance
$d{\cal B}(B\to X_s \mu^+\mu^-)/d\hat{s}$. However, as shown in
Fig.~\ref{FigDBAbfSCaseI}~(c), it nearly can't
reduce ${\cal B}(B\to X_s \mu^+\mu^-)$. So, if the future refined
experimental measurement on ${\cal B}(B\to X_s \mu^+\mu^-)$ is
significantly smaller than the SM prediction, S2 will be excluded
first. For  both S1 and S2 cases, the $Z^{\prime}$-induced effects on
$dA_{FB}(B\to X_s \mu^+\mu^-)/d\hat{s}$ are  tiny as shown in
Figs.~\ref{FigDBAbfSCaseI}~(b) and (d).

In our fitting, we find that the constraints on $B_{\mu\mu}^{L,R}$
are dominated by ${\cal B}^{(L,H)}(B\to X_s \mu^+\mu^-)$, while the
constraint from ${\cal B}(B_s\to \mu^+\mu^-)$ is very weak due to
the fact that there exits only upper bound at the moment. At the
quark level, since both $B\to X_s \mu^+\mu^-$ and $B_s\to
\mu^+\mu^-$ involve the same $b\to s \mu^+\mu^-$ transition, it is
interesting to see the $Z^{\prime}$ contributions to $B_s\to
\mu^+\mu^-$ within the parameter spaces constrained by $B\to X_s
\mu^+\mu^-$.

\begin{figure}[t]
\begin{center}
\epsfxsize=7cm \centerline{\epsffile{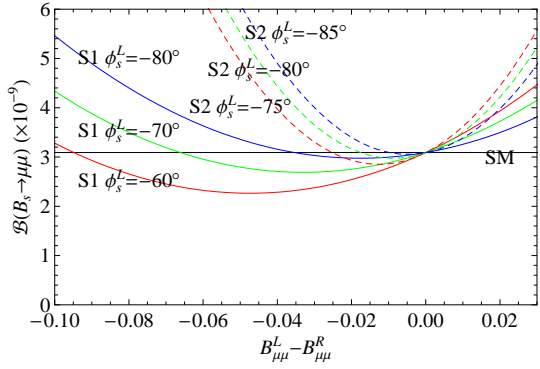}}
\centerline{\parbox{16cm}{\caption{\label{BsuuBuuLR}\small The
dependence of ${\cal B}(B_s\to \mu^+\mu^-)$ on
$B_{\mu\mu}^{L}-B_{\mu\mu}^{R}$ at
$\phi_s^L=-80^{\circ}$~($-85^{\circ}$),
$-70^{\circ}$~($-80^{\circ}$) and $-60^{\circ}$~($-75^{\circ}$) with
$|B_{sb}^{L}|=1.09\times10^{-3}$($2.20\times10^{-3}$) in S1~(S2).
}}}
\end{center}
\end{figure}

Fig.~\ref{BsuuBuuLR} shows the dependence of ${\cal B}(B_s\to
\mu^+\mu^-)$ on $B_{\mu\mu}^{L}-B_{\mu\mu}^{R}$ at different
$\phi_s^{L}$. In S1, with $B_{\mu\mu}^{R}=0$, we find that ${\cal
B}(B_s\to \mu^+\mu^-)$ is easier to be reduced with a smaller
$|\phi_s^{L}|$. Numerically, with $\phi_s^{L}\sim-65^{\circ}$ and
$B^{L}_{\mu\mu}\sim-0.05$, we get ${\cal B}(B_s\to
\mu^+\mu^-)=2.5\times10^{-9}$, which is $19\%$ smaller than the SM
prediction $3.1\times10^{-9}$. However, since
$|\phi_s^{L}({S2})|>|\phi_s^{L}({S1})|$, ${\cal B}(B_s\to
\mu^+\mu^-)$ is sensitive to seizable
$|B^{L}_{\mu\mu}-B^{R}_{\mu\mu}|$ and could be enhanced for most
parameter space of $\phi^{L}_{s}-|B^{L}_{\mu\mu}-B^{R}_{\mu\mu}|$.
 For S1~(S2), with $\phi_s^{L}=-80^{\circ}$~($-86^{\circ}$),
$B^{L}_{\mu\mu}=-0.06$($-0.02$), we find that the $Z^{\prime}$
contributions in case I could enhance ${\cal B}(B_s\to \mu^+\mu^-)$
by about $18\%$~($12\%$) compared with the SM prediction.

\begin{table}
 \begin{center}
 \caption{Numerical results for ${\cal B}(B\to\mu^+\mu^-)(\times10^{-9})$,
 ${\cal B}^{L,H}(B\to X_s\mu^+\mu^-)(\times10^{-7})$, and $A_{FB}(B\to
 X_s\mu^+\mu^-)(\times10^{-2})$.}
 \label{predictions}
 \vspace{0.5cm}
 \small
 \doublerulesep 0.7pt \tabcolsep 0.04in
 \begin{tabular}{lccccccccccc} \hline \hline
 Cases                         &\multicolumn{1}{c}{Exp.}&\multicolumn{1}{c}{SM} &\multicolumn{2}{c}{Case I} & \multicolumn{2}{c}{Case II} &\multicolumn{2}{c}{Case III} \\
                               &                       &                       & S1          & S2          & S1          &S2            & S1            & S2              \\\hline
 ${\cal B}(B_s\to\mu^+\mu^-)$      &$<47$                  &$3.1\pm0.5$        &$3.0\pm1.1$  &$3.3\pm0.8$  &$3.1\pm1.4$  &$3.5\pm1.1$   &$3.2\pm1.5$    &$3.5\pm1.1$            \\
 ${\cal B}(B\to X_s\mu^+\mu^-)$    &$43^{-12}_{+13}$       &$50\pm7$           &$46\pm10$    &$49\pm7$     &$50\pm7$     &$49\pm7$      &$46\pm11$       &$49\pm7$               \\
 ${\cal B}^{L}(B\to X_s\mu^+\mu^-)$&$16^{-4.8}_{+5.2}$     &$18\pm3.2$         &$17\pm4.3$   &$18\pm3.2$   &$18\pm3.1$   &$18\pm3.1$    &$16\pm4.5$     &$18\pm3.2$                       \\
 ${\cal B}^{H}(B\to X_s\mu^+\mu^-)$&$4.4\pm1.2$            &$4.5\pm0.6$        &$4.4\pm1.2$  &$4.4\pm1.2$  &$4.5\pm1.1$  &$4.4\pm1.1$   &$4.4\pm1.2$    &$4.4\pm1.1$                         \\
 $A_{FB}(B\to X_s\mu^+\mu^-)$      &---                    &$28\pm0.1$         &$29\pm2$     &$29\pm1.2$   &$22\pm6.6$   &$22\pm5.6$    &$18\pm13$       &$23\pm6.7$\\
 $A_{FB}^{L}(B\to X_s\mu^+\mu^-)$  &---                    &$0.5\pm0.3$        &$1.1\pm1.2$  &$1.0\pm0.8$  &$0.2\pm0.7$  &$0.2\pm0.5$   &$0.2\pm2.0$    &$0.4\pm0.9$\\
 $A_{FB}^{H}(B\to X_s\mu^+\mu^-)$  &---                    &$17\pm1.3$         &$17\pm1.4$   &$17\pm1.4$   &$14\pm4.4$   &$15\pm3.9$    &$11\pm7.3$     &$14\pm4.5$\\
 \hline \hline
 \end{tabular}
 \end{center}
 \end{table}

\begin{figure}[t]
\begin{center}
\epsfxsize=15cm \centerline{\epsffile{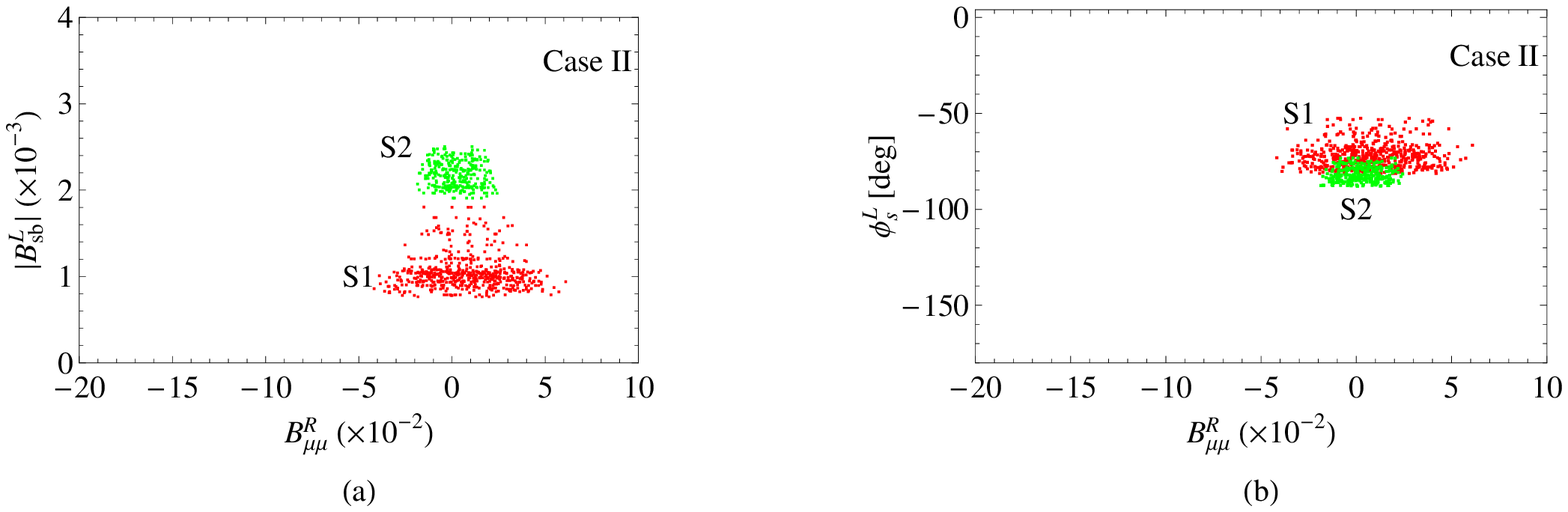}}
\centerline{\parbox{16cm}{\caption{\label{CaseII}\small The allowed
regions for the parameters $B_{\mu\mu}^{L}$ and $B_{\mu\mu}^{R}$.}}}
\end{center}
\end{figure}
\begin{figure}[t]
\begin{center}
\epsfxsize=15cm \centerline{\epsffile{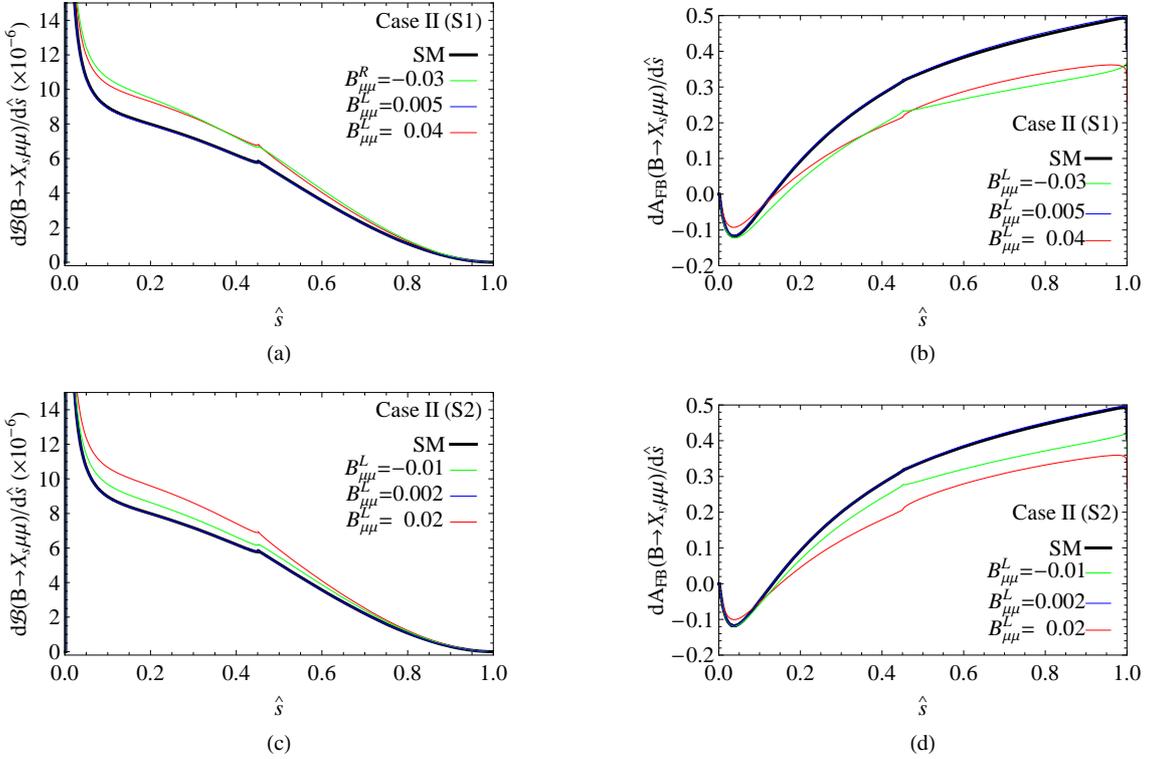}}
\centerline{\parbox{15cm}{\caption{\label{FigDBAbfSCaseII}\small The
dependence of $d{\cal B}(B\to X_s \mu^+\mu^-)/d\hat{s}$ and
$dA_{FB}(B\to X_s \mu^+\mu^-)/d\hat{s}$ on $\hat{s}$ with
$|B_{sb}^{L}|=1.09\times10^{-3}$($2.20\times10^{-3}$),
$\phi_s^L=-72^{\circ}$($-82^{\circ}$), and $B_{\mu\mu}^{R}=0$. The
other captions are the same as in Fig.~\ref{FigDBAbfSCaseI}.}}}
\end{center}
\end{figure}

\subsubsection*{Case~II: with $B^{R}_{\mu\mu}$ arbitrary, while taking $B_{\mu\mu}^{L}=0$.}
Taking $B^{L}_{\mu\mu}=0$, we are going to
evaluate the $Z^{\prime}$ effects induced by $B^{R}_{\mu\mu}$. The
allowed regions of the $Z^{\prime}$ parameters are shown in
Fig.~\ref{CaseII}. From Fig.~\ref{FigDBBuuLR}~(b), which shows the
dependence of ${\cal B}(B\to X_s \mu^+\mu^-)$ on $B^{R}_{\mu\mu}$,
one would observe that the minimal ${\cal B}(B\to X_s \mu^+\mu^-)$ corresponds
to the point $B^{R}_{\mu\mu}=0$. So, the $Z^{\prime}$ contributions
induced by $B^{R}_{\mu\mu}$ are nearly can't reduce ${\cal B}(B\to
X_s \mu^+\mu^-)$, which is confirmed by
Figs.~\ref{FigDBAbfSCaseII}~(a) and (c).

In Case~I, from Figs.~\ref{FigDBAbfSCaseI}~(b) and
(d), we find $dA_{FB}(B\to X_s
\mu^+\mu^-)/d\hat{s}$ is not very sensitive to $B^{L}_{\mu\mu}$.
However, comparing Figs.~\ref{FigDBAbfSCaseII}~(b) and
(d) with Figs.~\ref{FigDBAbfSCaseI}~(b)
and (d), we find $dA_{FB}(B\to X_s
\mu^+\mu^-)/d\hat{s}$ is very sensitive to the $Z^{\prime}$
contributions induced by $B^{R}_{\mu\mu}$. Since the $Z^{\prime}$
contributions to $E(\hat{s})$ are smaller than that to $D(\hat{s})$,
$A_{FB}(B\to X_s \mu^+\mu^-)$ can be reduced easily
rather than enhanced. With $B^{R}_{\mu\mu}=0.03$~($0.015$) and the central
values of the other theoretical parameters, $A_{FB}(B\to X_s
\mu^+\mu^-)$ could be reduced by a factor $12\%$~($9\%$) in S1~(S2)
compared to the SM prediction. At high/low $q^2$ region, it could be
reduced by  $12\%$~($13\%$)/$17\%$~($55\%$).

Taking $B^{L}_{\mu\mu}=0$, Fig.~\ref{BsuuBuuLR} shows the dependance
of ${\cal B}(B_s\to \mu^+\mu^-)$ on $B^{R}_{\mu\mu}$. At
$B^{R}_{\mu\mu}=0.03$ with $\phi_s^L-=65^{\circ}$, we get the small
${\cal B}(B_s\to \mu^+\mu^-)\sim2.5\times10^{-9}$, which is $18\%$
smaller than the SM prediction. With $B^{R}_{\mu\mu}=0.03$,
$\phi_s^L=-80^{\circ}$, ${\cal B}(B_s\to \mu^+\mu^-)$ is enhanced by
about $10\%$.

\begin{figure}[t]
\begin{center}
\epsfxsize=15cm \centerline{\epsffile{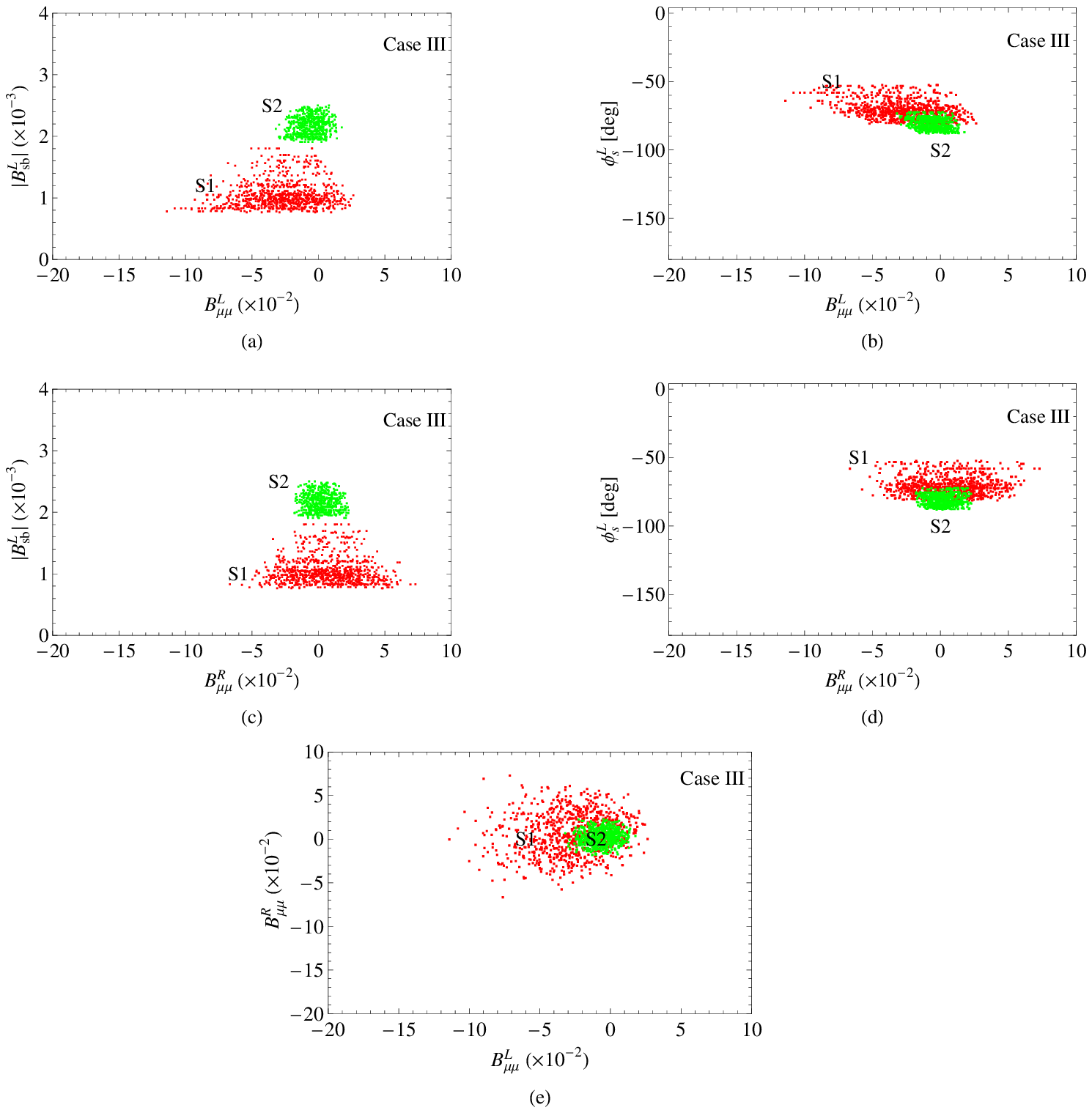}}
\centerline{\parbox{16cm}{\caption{\label{CaseIII}\small The allowed
regions for the parameters $B_{\mu\mu}^{L}$ and $B_{\mu\mu}^{R}$.}}}
\end{center}
\end{figure}

\begin{figure}[t]
\begin{center}
\epsfxsize=15cm \centerline{\epsffile{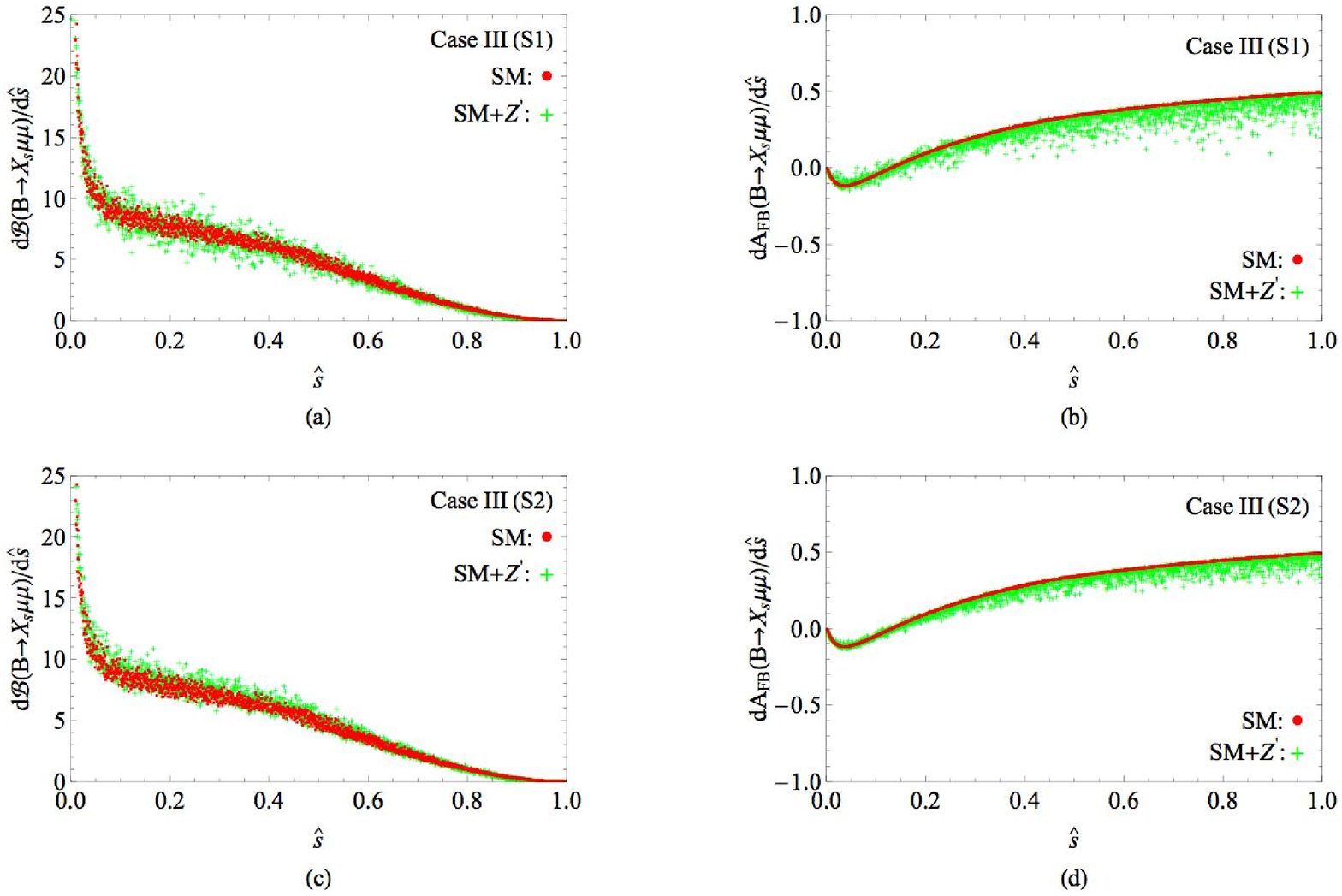}}
\centerline{\parbox{16cm}{\caption{\label{FigDBAbfSCaseIII}\small
The effects of the $Z^{\prime}$ contributions induced by
$B_{\mu\mu}^{L,R}$ on $d{\cal B}(B\to X_s \mu^+\mu^-)/d\hat{s}$ and
$dA_{FB}(B\to X_s \mu^+\mu^-)/d\hat{s}$. The other captions are the
same as in Fig.~\ref{FigDBAbfSCaseI}}}}
\end{center}
\end{figure}
\subsubsection*{Case~III: with both $B_{\mu\mu}^{L}$ and $B^{R}_{\mu\mu}$ arbitrary.}
More generally, we give up any assumptions for the $Z^{\prime}$
couplings $B_{\mu\mu}^{L}$ and $B^{R}_{\mu\mu}$. Because of the
interference effect between $B_{\mu\mu}^{L}$ and $B^{R}_{\mu\mu}$,
the allowed regions for these two parameters are now larger than the
ones in Case~I and Case~II, which are shown in Fig.~\ref{CaseIII}. From the figure, one can
find the correlations between the coupling parameters. Since $B^{L}_{sb}$ and its phase have been
constrained by $B_{s}-\bar{B}_{s}$ mixing,    $B_{\mu\mu}^{L}$ and $B^{R}_{\mu\mu}$ are
found to be small.  It is interesting to note that a model independent constraint on $C_{10}^{NP}$ and it
phase has been performed in Ref.\cite{Hiller} with $A_{FB}(B\to X_{s}\ell\ell)$. Combing our constraints
on   $B^{L}_{sb}$,  $B_{\mu\mu}^{L}$ and $B^{R}_{\mu\mu}$,  and re-scaling the combination by $V_{ts}$,
one can find our constraints are in good agreement with the magnitude of  $C_{10}^{NP}$ in Ref.\cite{Hiller}, but with
 a much stronger constraint on its phase.

The dependence of $d{\cal B}(B\to X_s \mu^+\mu^-)/d\hat{s}$ and
$dA_{FB}(B\to X_s \mu^+\mu^-)/d\hat{s}$ on $B_{\mu\mu}^{L}$ and
$B^{R}_{\mu\mu}$ has been discussed separately in the last two
cases. So, including all of the theoretical uncertainties, we just
present the dilepton invariant mass spectrum and the differential
normalized forward-backward asymmetry in
Fig.~\ref{FigDBAbfSCaseIII}. We find that in S1, as shown in
Fig.~\ref{FigDBAbfSCaseIII}~(a), ${\cal B}(B\to
X_s \mu^+\mu^-)$ could be either enhanced or reduced by $Z^{\prime}$
contributions. Since the $Z^{\prime}$ contributions induced by
$B^{R}_{\mu\mu}$ can hardly reduce ${\cal B}(B\to X_s \mu^+\mu^-)$
as discussed in Case~II, the $Z^{\prime}$ effects for reducing
${\cal B}(B\to X_s \mu^+\mu^-)$ are dominated by $B^{L}_{\mu\mu}$ as
discussed in Case~I, while the enhancement is due to both
$B^{L}_{\mu\mu}$ and $B^{R}_{\mu\mu}$.

However, in S2, as shown in Fig.~\ref{FigDBAbfSCaseIII}~(c), ${\cal B}(B\to X_s \mu^+\mu^-)$ can hardly be
reduced, which confirms our analysis in Case.~I and II. In both S1
and S2, as shown in Figs.~\ref{FigDBAbfSCaseIII}~(b)
and (d), $A_{FB}(B\to X_s \mu^+\mu^-)$ can be easily
reduced but hardly  be enhanced by $B^{R}_{\mu\mu}$.

With both $B^{L}_{\mu\mu}$ and $B^{R}_{\mu\mu}$ included, the
branching ratio for $B_s\to\mu^+\mu^-$ is affected by
$B_{\mu\mu}^{L}-B_{\mu\mu}^{R}$. Numerically, with
$B_{\mu\mu}^{L}-B_{\mu\mu}^{R}=-0.05$, $\phi^{L}_s=-65^{\circ}$ and
the central value of the other theoretical inputs, ${\cal
B}(B_s\to\mu^+\mu^-)$ is reduced by about $19\%$, which is the same
as in Case~I. With $B_{\mu\mu}^{L}=-0.05$, $B_{\mu\mu}^{R}=0.03$,
$\phi^{L}_s=-80^{\circ}$ and the central value of the other
theoretical inputs, we find that ${\cal
B}(B_s\to\mu^+\mu^-)=4.5\times10^{-9}$, which is about $46\%$ larger
than the SM prediction. So, if the coming measurements at LHCb and
super B factories present ${\cal B}(B_s\to\mu^+\mu^-)\sim10^{-8}$,
 the family non-universal $Z^{\prime}$ model will
suffer a serious challenge.

\section{Conclusion}
In conclusion, motivated by the observed $\bar{B}_s-B_s$ mixing
phase anomaly and the so-called ``$\pi K$ puzzle'', we have studied
a family non-universal $Z^{\prime}$ model to pursue possible
solutions. With the constrained $b-s-Z^{\prime}$ coupling by
$\bar{B}_s-B_s$ mixing and $B\to\pi K$ decays, we focus on the
$Z^{\prime}$ effects on the rare $B\to X_s \mu^+\mu^-$~(including
both the high and the low $q^2$ regions) and the purely leptonic
$B_s\to\mu^+\mu^-$ decays, both of which are also induced by FCNC
$b\to s$ transitions. Our main conclusions are summarized as:

\begin{itemize}
\item $\bar{B}_s-B_s$ mixing anomaly and
``$\pi K$ puzzle'' could be moderated simultaneously within such a
family non-universal $Z^{\prime}$ model. Corresponding to the two
fitting results S1 and S2 by UTfit collaboration, a new weak phase
$\phi^L_s\sim-72^{\circ}$ and $-82^{\circ}$ are crucial to resolve
these two problems.

\item Similar to the hierarchy of the CKM elements
$|V_{td}^{\ast}V_{tb}/V_{ts}^{\ast}V_{tb}|\sim0.2$, we find
 $|B_{db}/B_{sb}|\sim\mathcal {O}(10^{-1})$~($\lesssim
0.2$). So,  such a hierarchy should be hold within the
model. Our results also imply the relations
$B_{uu}^{L}<B_{dd}^{L}$ and $B_{uu}^{R}>B_{dd}^{R}$.

\item Combing  $B_{sb}^{L}$ restricted by $\bar{B}_s-B_s$ mixing and
$B\to\pi K$ decays, and $B_{\mu\mu}^{L,R}$ by   $B\to X_s \mu^+\mu^-$,
we find $B_{\mu\mu}^{L,R}\sim \mathcal
{O}(10^{-2})$. For observable ${\cal B}(B\to X_s \mu^+\mu^-)$, the
reduction effects is dominated by the $Z^{\prime}$ contributions
induced by $B_{\mu\mu}^{L}$ in S1. And, both the $Z^{\prime}$
contributions induced by $B_{\mu\mu}^{L}$ and $B_{\mu\mu}^{R}$ are
helpful to enhance it. The forward-backward symmetry $A_{FB}(B\to
X_s \mu^+\mu^-)$ is sensitive to the $Z^{\prime}$ contributions
induced by $B_{\mu\mu}^{R}$ but dull to the one induced by
$B_{\mu\mu}^{L}$.

\item With the strictly constrained $Z^{\prime}$ couplings  by
$\bar{B}_s-B_s$ mixing, $B\to\pi K$ and $B\to X_s \mu^+\mu^-$,
comparing with the SM prediction, we find ${\cal
B}(B_s\to\mu^+\mu^-)$ could be reduced/enhanced about $19\%$/$46\%$
by $Z^{\prime}$ contributions at most. The minimal value of ${\cal
B}(B_s\to\mu^+\mu^-)$ appears at the point
$B_{\mu\mu}^{L}-B_{\mu\mu}^{R}\sim-0.05$ with the minimal new weak
phase $\phi^{L}_s\sim-65^{\circ}$.
\end{itemize}

The  refined measurements for the (semi-)leptonic $B_{(s)}$
decay in the upcoming LHCb and super B factory will provide a
fertile testing ground for the SM and possible NP. Our analysis
about the $Z^{\prime}$ effects on the observables ${\cal
B}^{(H,L)}(B\to X_s \mu^+\mu^-)$, $A_{FB}^{(H,L)}(B\to X_s
\mu^+\mu^-)$ and ${\cal B}(B_s \to \mu^+\mu^-)$ are helpful to
confirm or refute the family non-universal $Z^{\prime}$ model.

\section*{Acknowledgments}
X.~Q.~Li acknowledges support from the Alexander-von-Humboldt
Foundation. The work is supported by the National Science Foundation under contract
Nos.10675039 and 10735080.
\begin{appendix}

\section*{Appendix~A: Theoretical input parameters}
For the CKM matrix elements, we adopt the fitting results from UTfit
collaboration~\cite{UTfit,UTfitCKM}
\begin{eqnarray}
&&\overline{\rho}=0.154\pm0.022\,(0.177\pm0.044), \qquad
\overline{\eta}=0.342\pm0.014\,(0.360\pm0.031),\nonumber\\
&&|V_{td}/V_{ts}|=0.209\pm0.0075\,(0.206\pm0.012),\nonumber\\
&&|V_{cb}|=(4.13\pm0.05)\times10^{-2}\,((4.12\pm0.05)\times10^{-2}),
\end{eqnarray}
with $\overline{\rho}=\rho\,(1-\frac{\lambda^2}{2})$ and
$\bar{\eta}=\eta\,(1-\frac{\lambda^2}{2})$. The values given in the
brackets are the CKM parameters in presence of generic NP, and used
in our calculation when the $Z^{\prime}$ contributions are included.

As for the quark masses, there are two different classes appearing
in our calculation. One type is the current quark mass which is
scale dependent. Here we take
\begin{eqnarray}
\frac{\overline{m}_s(\mu)}{\overline{m}_q(\mu)}&=&27.4\pm0.4\,~\cite{HPQCD:2006},\quad
\overline{m}_{s}(2\,{\rm GeV}) =87\pm6\,{\rm
MeV}\,~\cite{HPQCD:2006},
\quad\overline{m}_{c}(\overline{m}_{c})=1.27^{+0.07}_{-0.11}\,{\rm
GeV}~\cite{PDG08}\,,\nonumber\\
\overline{m}_{b}(\overline{m}_{b})&=&4.20^{+0.17}_{-0.07}\,{\rm
GeV}~\cite{PDG08}\,,\quad
\overline{m}_{t}(\overline{m}_{t})=164.8\pm1.2\,{\rm
GeV}~\cite{PDG08}\,,
\end{eqnarray}
where $\overline{m}_q(\mu)=(\overline{m}_u+\overline{m}_d)(\mu)/2$,
and the difference between $u$ and $d$ quark is not distinguished.
The other one is the pole quark mass. In this paper, we
take~\cite{PDG08,PMass}
\begin{eqnarray}
 &&m_u=m_d=m_s=0, \quad m_c=1.61^{+0.08}_{-0.12}\,{\rm GeV},\nonumber\\
 &&m_b=4.79^{+0.19}_{-0.08}\,{\rm GeV}, \quad m_t=172.4\pm1.22\,{\rm GeV}.
\end{eqnarray}

As for the B-meson lifetimes and decay constants, we
take~\cite{PDG08,DecayCon}
\begin{eqnarray}
 &&\tau_{B_{u}} = 1.638\,{\rm ps}\,, \qquad  \tau_{B_{d}}=1.530\,{\rm ps}\,,\\
 &&f_{B_{u,d}}=(190\pm13)~{\rm MeV}\,, \qquad
 \sqrt{\hat{B}_{B_{d}}}f_{B_{d}}=(216\pm15)~{\rm MeV}\,,\nonumber\\
 &&f_{B_{s}}=(231\pm15)~{\rm MeV}\,, \qquad
 \sqrt{\hat{B}_{B_{s}}}f_{B_{s}}=(266\pm18)~{\rm MeV}\,.
\end{eqnarray}

\section*{Appendix~B: Derivation for the Eq.~(\ref{BrBsmumu})}
From the effective Hamiltonian for $B_s\to l^+l^-$ decay given by
Eqs.~(\ref{SMHbsll}) and (\ref{ZPHbsll}), the amplitude for the
$B_s\to l^+l^-$ decay can be written as
\begin{eqnarray}
 A&=&A_{SM}+A_{Z^{\prime}}\,,\\
 A_{SM}&=&-\frac{G_{F}}{\sqrt{2}}V_{tb}V_{ts}^{\ast}\frac{\alpha}{2\pi
 \sin^2\theta_W}Y(x_t)\langle
 \mu^+\mu^-|\bar{s}\gamma^{\mu}(1-\gamma_5)b\otimes\bar{\mu}\gamma_{\mu}(1-\gamma_5)\mu|\bar B_s\rangle\,,\\
 A_{Z^{\prime}}&=&-\frac{G_{F}}{\sqrt{2}}V_{tb}V_{ts}^{\ast}
 \left[ \frac{-2B_{sb}^{L}B_{\mu\mu}^L}{V_{tb}V_{ts}^{\ast}}\langle
 \mu^+\mu^-|\bar{s}\gamma^{\mu}(1-\gamma_5)b\otimes\bar{\mu}\gamma_{\mu}(1-\gamma_5)\mu|\bar B_s\rangle
 \right. \nonumber\\
 &&
 \left. +\frac{-2B_{sb}^{L}B_{\mu\mu}^R}{V_{tb}V_{ts}^{\ast}}\langle
 \mu^+\mu^-|\bar{s}\gamma^{\mu}(1-\gamma_5)b\otimes\bar{\mu}\gamma_{\mu}(1+\gamma_5)\mu|\bar B_s\rangle
 \right].
\end{eqnarray}
After parameterizing the hadron parts, the above equations could be
rewritten as
\begin{eqnarray}
A_{SM}&=&\frac{G_{F}}{\sqrt{2}}V_{tb}V_{ts}^{\ast}if_{B_s}P_{B_s}^{\mu}\frac{\alpha}{2\pi
 \sin^2\theta_W}Y(x_t)\bar{\mu}\gamma_{\mu}(1-\gamma_5)\mu\,,\\
A_{Z^{\prime}}&=&\frac{G_{F}}{\sqrt{2}}V_{tb}V_{ts}^{\ast}if_{B_s}P_{B_s}^{\mu}
\left[
\frac{-2B_{sb}^{L}B_{\mu\mu}^L}{V_{tb}V_{ts}^{\ast}}\bar{\mu}\gamma_{\mu}(1-\gamma_5)\mu+\frac{-2B_{sb}^{L}B_{\mu\mu}^R}{V_{tb}V_{ts}^{\ast}}\bar{\mu}\gamma_{\mu}(1+\gamma_5)\mu
\right],
 \end{eqnarray}
where we have defined $\langle0|\bar{s}\gamma^{\mu}(1-\gamma_5)b|\bar B_{s}\rangle=-if_{B_s}P_{B_s}^{\mu}$, with $f_{B_s}$ being the $B_s$-meson decay constant.
For simplicity, we introduce
\begin{align}
A_1&\equiv\bar{\mu}\pslash_{B_s}(1-\gamma_5)\mu\,, &
A_2&\equiv\bar{\mu}\pslash_{B_s}(1+\gamma_5)\mu\,, \nonumber\\
B&\equiv\frac{G_{F}}{\sqrt{2}}V_{tb}V_{ts}^{\ast}if_{B_s}\,, &
C&\equiv \frac{\alpha}{2\pi\sin^2\theta_W}Y(x_t)\,, \nonumber \\
D_1&\equiv\frac{-2B_{sb}^{L}B_{\mu\mu}^L}{V_{tb}V_{ts}^{\ast}}\,, &
D_2&\equiv \frac{-2B_{sb}^{L}B_{\mu\mu}^R}{V_{tb}V_{ts}^{\ast}}\,.
\end{align}
Then the total decay amplitude can be written as
\begin{eqnarray}
A&=&B[(C+D_1)A_1+D_2A_2],  \\
|A|^2&=&|B|^2\left[|C+D_1|^2|A_1|^2+|D_2|^2|A_2|^2+(C+D_1)^{\ast}D_2A_1^{\ast}A_2
+(C+D_1)D_2^{\ast}A_1A_2^{\ast}\right]\,.
\end{eqnarray}
It is easy to get
\begin{eqnarray}
|A_1|^2&=&|A_2|^2=8m_{\mu}^2m_{B_s}^2\,,\\
A_1^{\ast}A_2&=&A_1A_2^{\ast}=-8m_{\mu}^2m_{B_s}^2\,.
\end{eqnarray}
So,
\begin{eqnarray}
|A|^2&=&|B|^{2} 8m_{\mu}^2 m_{B_s}^2 |(C+D_1)-D_2|^2\,,\nonumber\\
&=&\frac{G_F^2}{2}|V_{tb}V_{ts}^{\ast}|^{2} 8m_{\mu}^2m_{B_s}^2
\left| \frac{\alpha}{2\pi\sin^2\theta_W}Y(x_t)-\frac{2B_{sb}^{L}(B_{\mu\mu}^{L}
-B_{\mu\mu}^{R})}{V_{tb}V_{ts}^{\ast}}
\right|^{2}.
\end{eqnarray}
Finally, with $|P_c|=\frac{1}{2}\sqrt{m_{B_s}^2-4m_{\mu}^2}$, we get
\begin{eqnarray}
 {\cal B}(B_s\to\mu^+\mu^-)&=&\tau_{B_s}\frac{|P_c|}{8\pi m_{B_s}^2}|A|^2\nonumber\\
 &=&\tau_{B_s}\frac{G_F^2}{4\pi}f_{B_s}^2m_{\mu}^2m_{B_s}
 \sqrt{1-\frac{4m_{\mu}^2}{m_{B_s}^2}}|V_{tb}V^{\ast}_{ts}|^2\nonumber\\
                        & &\times  \left| \frac{\alpha}{2\pi \sin^2\theta_W}Y(x_t)
 -2\frac{B_{sb}^{L}(B_{\mu\mu}^L-B_{\mu\mu}^R)}{V_{tb}V^{\ast}_{ts}}\right|^2\,.
\end{eqnarray}
\end{appendix}

\end{document}